\documentclass[preprint,12pt]{elsarticle}

\usepackage{amssymb}
\usepackage{amsmath}
\usepackage{subcaption}
\usepackage{miller}
\usepackage{hyperref}

\journal{Journal of the Mechanics and Physics of Solids}

\begin{document}

\begin{frontmatter}



\title{Femtosecond two-pulse laser approach for spall failure in thin foils}


\author[inst1]{Mewael Isiet}

\affiliation[inst1]{organization={Department of Mechanical Engineering},
            addressline={The University of British Columbia}, 
            postcode={Vancouver}, 
            state={British Columbia },
            country={V6T 1Z4, Canada}}

\author[inst2,inst3]{Yunhuan Xiao}
\author[inst2,inst3]{Jerry I. Dadap}
\author[inst2,inst3]{Ziliang Ye}
\author[inst1]{Mauricio Ponga}

\affiliation[inst2]{organization={Department of Physics \& Astronomy},
            addressline={The University of British Columbia}, 
            city={Vancouver},
            postcode={British Columbia}, 
            country={V6T 1Z1, Canada}}

\affiliation[inst3]{organization={Quantum Matter Institute},
	addressline={The University of British Columbia}, 
	city={Vancouver},
	postcode={British Columbia}, 
	country={V6T 1Z1, Canada}}
	
\begin{abstract}
Spall failure in materials occurs when tensile waves, propagating through a material, interact leading to failure once the generated hydrostatic stress exceeds the material's strength. 
In this study, we introduce a novel two-pulse laser approach to induce spall failure by simultaneously illuminating both free surfaces of micro- and nanoscale-thick Ni foils using femtosecond laser pulses.
By comparing this method with the conventional single-pulse approach, we demonstrate the unique effects of introducing a secondary laser pulse.
Experimental observations utilizing electron microscopy and focused ion beam milling reveal that the proposed approach effectively induces failure at the center of the sample, resulting from the interaction of unloading tensile waves in the bulk. 
Notably, the spall threshold for nanoscale Ni foils is significantly reduced in the two-pulse approach compared to the single-pulse method. 
While the single-pulse method requires a threshold fluence of 2,500 $\mathrm{J \cdot m^{-2}}$, the two-pulse approach lowers this to 1,750 $\mathrm{J \cdot m^{-2}}$, highlighting its potential to generate higher hydrostatic stress with less laser pulse energy.
Both experimental results and molecular dynamics simulations indicate ductile failure mechanisms involving nucleation, growth, and coalescence of voids, preceded by the emission of stacking faults and Shockley partial dislocations. 
These findings suggest that the two-pulse approach can be an effective approach for high-throughput photomechanical spallation experiments in metals.
\end{abstract}

\begin{highlights}
\item Two-pulse laser approach shifts spall failure away from free surfaces.
\item Incorporating a secondary pulse lowers laser fluence required.
\item MD simulations show spall failure from hydrostatic stress via tensile wave interaction.
\end{highlights}

\begin{keyword}
Laser-induced spallation\sep Shockwaves\sep  Dynamic failure\sep  Nickel
\end{keyword}

\end{frontmatter}


\section{Introduction}
Failure of materials under shock loading is an essential field of study with applications in defense, high-energy impacts, and planetary collision. 
When two or more objects collide, a series of shock waves propagate in materials, causing significant microstructural and temperature changes, and in extreme cases, it could lead to phase transformation \cite{ramesh2015review}.
When these shock waves reflect at the material's free surfaces, the traction-free boundary condition requires that compressive shock waves reflect-with the same magnitude but opposite displacement-to satisfy the traction-free condition. 
Interestingly, when two tensile waves collide under uniaxial strain loading, they produce a stress state characterized by large hydrostatic tensile stress \cite{meyers1983dynamic}.
This hydrostatic stress is the primary cause of spall failure in materials under shock loading, driven by nanovoid nucleation, growth, and coalescence. 
Spall failure is, in essence, a dynamic ductile-type failure and plays an essential role in determining the mechanical resistance of materials under high-velocity impacts and ballistic applications.

Spall is a catastrophic failure mode that could compromise structural components \cite{righi2021towards}.
Due to its central role in ballistic applications, spall failure has been an essential subject of study in structural, polymeric, and biomedical materials \cite{xie2018shock,hagerman2007evaluation,strachan2001critical,farbaniec2017spall}. 
The primary method for characterizing spall strength is through plate impact experiments \cite{williams2012spall}.
In plate impact experiments, a flyer plate is accelerated to high-velocities (typically between 100 to 600 m$\cdot$s$^{-1}$) and impacted onto a target plate made of the material one wants to test.
Typical flyer plate sizes between 1 to 5 mm in thickness and 10 to 100  mm in diameter \cite{zhang2022shock, ye2019spallation, xie2018shock} where the flyer plate thickness is one-half the target's thickness to maximize tensile damage at the centre of the target \cite{gray2018structure}.
Due to the thickness of the projectile, plate impact experiments produce shock loads that sample the microstructure of the materials to find weak spots that would result in spall failure of the material. 
A drawback of the plate impact experiment is the large amount of energy required to accelerate the projectile, which has significant mass. 
This issue becomes even more critical at high impact velocities and strain rates, drastically limiting the number of available facilities worldwide.
Other drawbacks are proper alignment of the projectile target and sensing free-surface velocity (which is mitigated using Photon Doppler Velocimeters (PDVs)) \cite{ramesh2008high}.
Regardless of these drawbacks, plate impact experiments provide the ultimate validation step for materials and devices before they can be subject to ballistic loads.

However, the need for faster, cheaper, and higher-throughput techniques for spall failure characterization has motivated the development of novel experimental techniques to complement the plate impact experiment.
For instance, researchers have recently developed laser-driven micro-flyer impact facilities, where the plate impact experiment conditions are reproduced but on much smaller scales (a typical flyer plate is about 0.05 mm thick and 1 mm in diameter) \cite{brown2012simplified, mallick2017investigating}.
By reducing the size of the flyer plate, the kinetic energy needed to accelerate it is much smaller and can be conveniently provided by powerful lasers. 
Additionally, by reducing the sizes of the specimens, very high-strain rates can be achieved, which may be challenging in traditional plate impact experiments. 
Several laser-driven flyer plate facilities have been developed with remarkable success and have been used to predict the spall strength of various materials, including Mg \cite{mallick2019laser}, Al \cite{PAISLEY1992825}, Au \cite{robbins2000laser}, Cu \cite{warnes1996hugoniot}, Ni \cite{wang2017laser} and boron carbide \cite{mallick2020dynamic}.

Another technique developed in the early 1960s is the laser spall experiment \cite{askaryon1963use,Neuman1964}. 
In this setup, a powerful laser is used to illuminate one of the traction-free surfaces of a specimen (target) \cite{de2018laser}. 
The photons interact with the electrons in the target material, which is sometimes coated with an absorption layer to enhance energy deposition.
The absorption layer, typically a thin ($\sim$20 nm) noble metal (Au, Ag, or Cu) with a high plasmonic response to the laser wavelength, aids in this process, though other materials have also been used \cite{ehsani2021evolution}.
The high-intensity single laser pulse generates a sharp temperature increase in the loaded surface, producing rapidly expanding plasma and compressive waves via the rocket effect \cite{remington2018spall}. 
With the shorter pulse durations (from femtoseconds to nanoseconds) and large energy deposition capabilities of lasers, the rapid interaction between photons, electrons, and phonons lead to a rapid energy exchange, resulting in  high-strain rate loads that exceed those in traditional plate impact experiments. 
Furthermore, under laser-shock compression, self-quenching occurs as the temperature reaches equilibrium in milliseconds, whereas in plate impact experiments, this time is on the order of seconds \cite{meyers2008deformation,cao2005effect}. 
Thus, it is often assumed that post-laser-shocked materials yield microstructures closer to the ones generated at the shock front, allowing for a more accurate post-mortem characterization \cite{remington2017deformation}.

Due to these benefits and recent advances in laser technology, laser shock experiments have become a viable and appealing option for studying spall failure in materials. 
However, it has been shown in atomistic simulations \cite{leveugle2004photomechanical, abou2018spallation, gill2011ultrashort, zhou2022efficient} that bulk metals, under single-pulse laser shock loads at the fluence threshold for spall failure, may fail when the tensile component of the pressure wave passes through the melted region, i.e., near the loaded surface where the activation energy for nucleation reaches a minimum value \cite{shugaev2019thermodynamic}.
Due to the shorter pulse duration in single-pulse laser shock experiments, spall could occur very close to the rear surface  \cite{gill2011ultrashort}, as demonstrated in numerous experimental works \cite{gilath1988laser, kingstedt2015ultra, hu2009high, eliezer1990laser}.
Therefore, the conventional laser-based spall experiments may induce void-surface interactions, which could influence the mechanical response of the material \cite{crone2015capturing, leu2021effects}, or involve processes such as micro-jetting or micro-spalling, leading to the ejection of thin jets or droplets from the unloaded free surface \cite{de2018picosecond}. 
These effects are heavily influenced by pulse duration, laser intensity, and surface roughness \cite{resseguier2010dynamic}.

In this work, we propose an alternative approach to induce spall failure using two laser pulses to mimic the conditions imposed in plate impact experiments. 
The use of femtosecond laser pulses is particularly advantageous as it ensures that laser heating time does not exceed the electron-phonon coupling time, which minimizes heat diffusion and allows sufficient pressure development.
It has been previously demonstrated that femtosecond lasers can effectively control laser-induced phenomena such as melting, spallation, and phase transitions \cite{Downer1985, Reitze1989, Unger2012, CuqLelandais2009} by controlling the laser fluence. 
We investigate the shock wave behavior in nickel (Ni) using a combination of experimental observations, theoretical modeling and atomistic simulations. 
In the proposed setup, a single femtosecond laser beam is split into two laser beams illuminating both the target's front and back surfaces. 
These two laser beams generate a rapid temperature and pressure increase at these surfaces and propagate two shock loads, containing a compression front followed by an unloading tensile wave, traveling in opposite directions. 
The interaction of the unloading tensile waves in the middle of the specimen develops a purely hydrostatic tensile stress at the center of the specimen. 
Through experiments, we first demonstrate the spall behavior induced by the proposed two-pulse approach and compare it to the conventional single-pulse laser method. 
Our results show that failure can occur deeper within the material, far from the free surfaces, when the resultant tensile stress exceeds the spall strength.
Additionally, the two-pulse method reduces the laser fluence threshold required for spall, mitigating potential thermal effects.
After establishing the effectiveness of this approach experimentally, we further investigate the shock and spall behavior, and failure mechanisms using molecular dynamics simulations.
\section{Methodology}
\subsection{Experimental setup}
\begin{figure}[h!]
	\centering
	\includegraphics[width=1.0\textwidth]{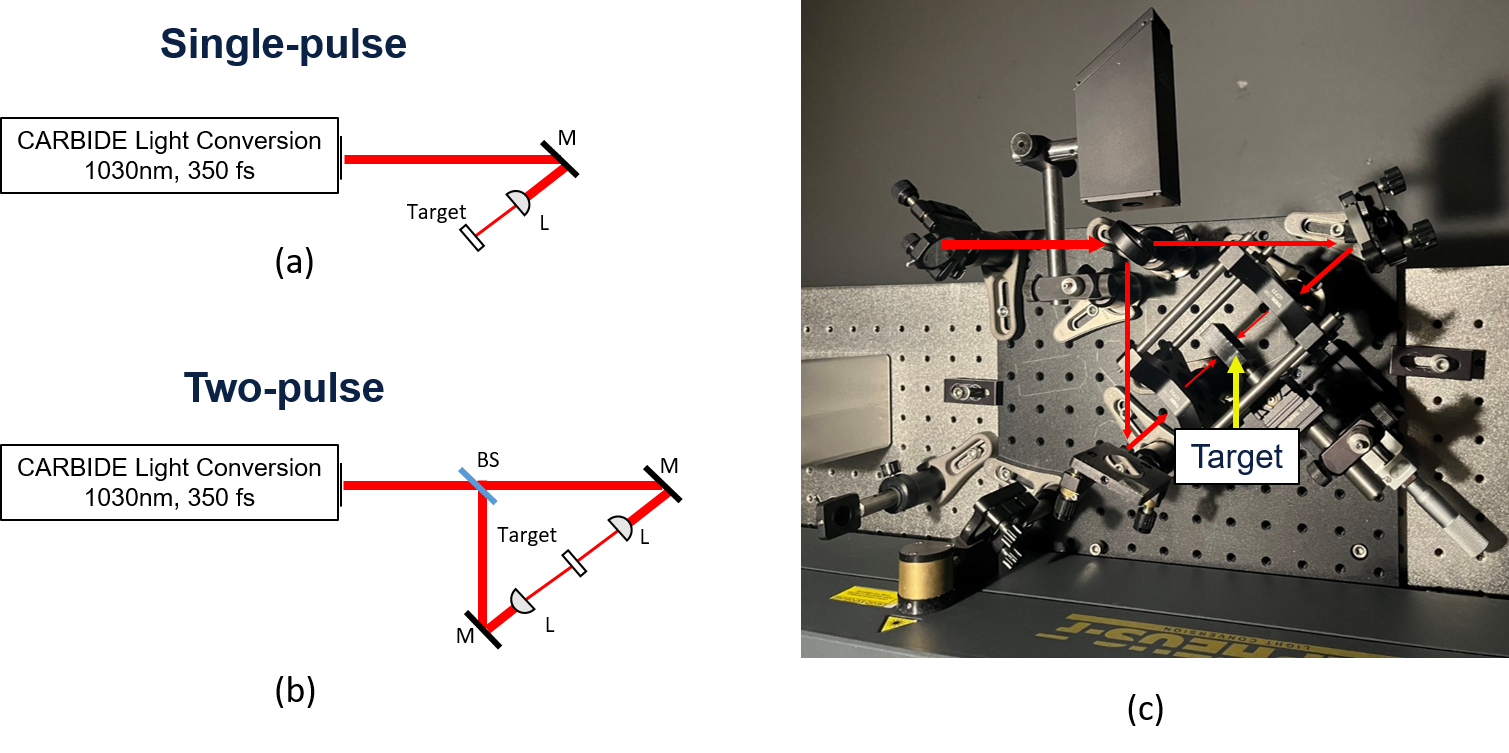}
	\caption{Schematic of the (a) single- (illumination of the front surface only) and (b) two-pulse (illumination of the front and back surfaces) laser-induced spallation setup. (c) Setup for the single and two-pulse experiments. M: mirror, L: lens, BS: beam splitter.}
	\label{fig:experimental_layout}
\end{figure}
Figure \ref{fig:experimental_layout} illustrates the layout of the laser-spallation setup, where (a) in the single-pulse approach, the laser beam is directed to a single surface, and (b) in the two-pulse configuration, a beamsplitter is used to split the laser beam into two, illuminating both surfaces of the sample. Figure \ref{fig:experimental_layout}(c) shows the setup for the single and two-pulse experiments. 
The laser-induced spall experiments were performed under ambient conditions using an ultra-short solid-state laser source (Carbide CB3-40W, Light Conversion) with a maximum measured pulse energy of 60 $\mu$J.
The laser source emitted pulses with a full width half maximum (FWHM) beam diameter of 3.9 $\pm$ 0.4 mm and a pulse duration of 350 fs at a wavelength $\lambda$ = 1030 $\pm$ 10 nm.
The emission of single pulses was controlled using a pulse-picker integrated into the master laser system, ensuring precise selection and emission of individual laser pulses.
To focus the laser, a pair of plano-convex lenses with a 40 mm focal length (LA1304-AB, Thorlabs) was employed, which allowed us to vary the spot size by adjusting the sample-to-lens distance.
The smallest laser beam spot size, approximately 35  $\mu$m, was measured using the knife-edge method \cite{Mylonakis2018}, resulting in laser fluences ranging from 5,000 to 25,000 $\mathrm{J \cdot m^{-2}}$. 
By increasing the spot size to approximately 80  $\mu$m, we achieved a lower range of laser fluences, from 850 to 5,000 $\mathrm{J \cdot m^{-2}}$. 
Laser fluence was carefully adjusted by varying the pulse energy via a power attenuator. 
To measure the energy of individual pulses, the laser was operated at a relatively low repetition rate of 200 Hz. 
The average laser power was measured using a calibrated powermeter (S130C, Thorlabs) placed directly before the sample.
Other optical elements included high-reflectivity mirrors (PF10-03-P01, Thorlabs) to align the beam, and a 50:50 beamsplitter plate (BSW11, Thorlabs) to split the laser beam into two equal parts, ensuring equal laser fluences on both surfaces in the two-pulse configuration.

Table \ref{tbl:1} provides a detailed overview of the current \emph{state-of-the-art} applications of the single-pulse laser approach for inducing spallation. 
A key limitation noted in the literature \cite{Zhigilei2009,shugaev2021laser} is that the use of extremely high laser fluences often triggers additional thermal phenomena, such as plasma formation, phase explosions, and surface boiling, which can complicate and influence the spall failure \cite{Tien1999}. 
Furthermore, the lack of stress confinement when using nanosecond laser pulses \cite{leveugle2004photomechanical, Demaske2010} results in limited pressure build-up, reducing the potential for effective damage initiation. 
As a result, these conventional nanosecond single-pulse laser methods typically require significantly higher laser fluences to induce spall failure, which unfortunately introduces undesirable thermal effects.
In contrast, the single- and two-pulse femtosecond laser approaches presented in this study achieve spall failure with much lower energy requirements. 
Since spallation is predominantly a mechanically-induced failure, driven by the interaction of tensile waves, the use of a femtosecond laser is highly suitable for these experiments, minimizing thermal influences and maximizing the potential of hydrostatic stress generation.
\begin{table}[h!]
	\scriptsize
	\begin{center}
		\caption{Summary of the state-of-the-art application of the conventional single-pulse laser setup in measuring the spall strength. sc: single crystalline, pc: poly crystal, nc: nano crystalline, PS: Polystyrene, HEA: high entropy alloy.}\label{tbl:1}    
		\begin{tabular}{p{1.0cm} p{1.5cm} p{1.4cm} p{1.2cm} p{1.2cm} p{1.5cm} p{2.0cm} p{.4cm}}
			\hline
			\hline
			Target & Absorption layer & Thickness (mm) & Pulse energy (J) & Spot size (mm) & Pulse duration (ns) & Spall strength (GPa) & Ref. \\ \hline     \hline
			Sn (sc) & -- & 2.0 -- 4.0 & 22 -- 990 & 1.8 -- 4.0 & 3.0 -- 5.0 & 0.7 -- 1.7 & \cite{de2007spallation} \\
			Fe (sc) & PS (30 $\mu$m) & 0.1 -- 0.17 & -- & 1.0 & 10.0 & 9.9 -- 10.7 & \cite{righi2021towards} \\
			Fe (pc) & PS (30 $\mu$m) & 0.1 -- 0.25 & -- & 1.0 & 10.0 & 4.3 -- 8.5 & \cite{righi2021towards} \\
			Fe (nc) & PS (30 $\mu$m) & 0.1 -- 0.2 & -- & 1.0 & 10.0 & 4.9 -- 6.6 & \cite{righi2021towards} \\
			Au (sc) & -- & 0.3 & 4.4\textit{e}03 & -- & 3.0 & 3.1 -- 5.6 & \cite{lescoute2009spallation} \\
			Ta (sc) & PS (20 $\mu$m) & 0.05 -- 0.25 & 100 & -- & -- & 9.0 -- 11.3 & \cite{remington2018spall} \\
			Ta (pc) & PS (20 $\mu$m) & 0.25 & 100 & -- & -- & 8.2 & \cite{remington2018spall} \\
			Ta (nc) & PS (20 $\mu$m) & 0.25 & 100 & -- & -- & 7.0 & \cite{remington2018spall} \\
			HEA & PS (20 $\mu$m) & 0.1 & 100 -- 200 & 3.0 & 1.0 & 8.0 & \cite{thurmer2022exceptionally} \\
			V (sc) & -- & 0.25 & 106 -- 423 & -- & 3.0 -- 8.0 & 5.0 - 9.0 & \cite{Jarmakani2010} \\
			Mg (sc) & -- & 0.07 -- 0.2 & 6.0 -- 17.0 & 1.0 & 4.0 & 1.8 -- 2.0 & \cite{de2017spall} \\
			W (sc) & Al (0.4 $\mu$m) & 0.2 -- 2.0 & 0 -- 0.3 & 1.0 -- 2.0 & 5.0 & 2.5 -- 3.4 & \cite{hu2009high} \\
			Cu (sc) & Al (0.4 $\mu$m) & 0.5 & 10 -- 80 & 0.2 -- 1.0 & 2.0 -- 5.0 & 3.0 -- 9.0 & \cite{moshe1998increase} \\
			Al (pc) & -- & 0.05 -- 0.2 & 30 & 4.0 & 3.0\textit{e}-4 & 1.5 -- 3.9 & \cite{CuqLelandais2009} \\
			Ni (pc) & -- & 0.0025 & 30\textit{e}-6 & 50\textit{e}-3 & 3.5\textit{e}-4 & 12.9 & Ours \\
			\hline
			\hline
		\end{tabular}    
	\end{center}
\end{table}

In this work, microscale- and nanoscale-thick Nickel foils with a thickness of 2.5 $\mu$m and 200 nm, respectively, were used as target materials to compare the single- and two-pulse laser approaches.
The nanoscale Ni foil was deposited onto a glass substrate, while the microscale (free-standing) Ni foil was affixed to a sample holder for integration into the experimental setup.
Commercially available microscale Ni foil (Thermo Fisher Scientific) samples with 99.95$\%$ purity were used for this spall study.
Using an XY stage, it was possible to systematically illuminate multiple spots across the surface of the foils.
For the single-pulse approach, the nanoscale foil was irradiated both directly and through the glass substrate, whereas in the two-pulse approach one surface was irradiated directly and the other through the glass substrate.
The microscale foil was irradiated directly in both approaches.
Post-mortem microstructural analysis was conducted using scanning electron microscopy (SEM, TESCAN AMBER) for both types of Ni foils.
The SEM was operated at an accelerating voltage of 2 keV and a beam current of 100 pA with a working distance of approximately 10 mm, which provided high-resolution images of the surface morphology and spallation features.
Additionally, the sub-surface of the irradiated microscale-thick Ni foil was exposed using a focused ion beam (FIB, TESCAN AMBER) operated at  30 keV and beam currents ranging from 1 to 30 nA, to perform precise milling and cross-sectional analysis of the spallation region. 
Prior to the FIB-milling, a thin protective layer of platinum was deposited on the surface of the illuminated foil to preserve the internal structure and prepare a clean and sharp FIB cut.
This approach enabled us to reveal internal damage, including void nucleation and coalescence, by sequentially removing material from the sample's surface layer by layer.
It should be noted that the objective of these experiments was to qualitatively describe the effects of the two-pulse laser approach on spall behavior and compare the proposed approach with the single-pulse approach.
\subsection{Atomistic simulation of laser-metal interaction}
In metallic materials, the energy provided by a laser source is absorbed by the electrons near the free surface whose energy is close to the Fermi level. 
This energy is then rapidly-within femtoseconds-equilibrated among the surrounding electrons and is eventually transferred to the phonons.
The energy transfer between electrons and phonons is controlled by the strength of a material's electron-phonon coupling ($G$). 
When the laser pulse duration is comparable to or shorter than the time required for electron-phonon thermalization, a state of thermal nonequilibrium is established, where the electron and lattice temperatures differ \cite{ivanov2003combined}, leading to complex heat exchanges between these subsystems.
There is also a condition for stress confinement dependent on the material's optical absorption depth ($L_p$) and bulk sound speed ($c_0$) \cite{zhigilei2000microscopic}.

Classical molecular dynamics (MD) has been employed to study laser-metal interactions \cite{zhidkov2001short, leveugle2004photomechanical, hakkinen1993superheating}.
However, because MD does not account for the electronic contribution to thermal conductivity, it leads to a nonphysical description of the mechanical response of metals under laser irradiation.
To remedy this issue, the two-temperature model has been employed to couple the electronic subsystem to the phonons in MD \cite{kagnov1957relaxation,anisimov1974electron}. 
In this work, we employ a variation of this method, called the local two-temperature model ($\ell$2T-MD) model \cite{ponga2018unified, ullah2019new}, which uses the Fokker-Planck equation, implemented via a master equation, to simulate heat conduction in the electronic subsystem.
The main advantage of this approach is that it can be seamlessly integrated with MD without introducing any basis set to compute Fourier's law. 
Next, we briefly summarize the $\ell$2T-MD model and the modifications to simulate the light-matter interaction.

Consider a system with $N$ atoms and assume that each atom is assigned with a variable electronic temperature, $T_{i}^{e}$, for each atom $i$.
We then parametrize the electron temperature as a dimensionless quantity using an arbitrary maximum temperature value  $T_\mathrm{max}^{e}$. 
The dimensionless temperature field is defined as $\theta_{i}^{e} = \frac{T_{i}^{e}}{T_\mathrm{max}^{e}}$ and ranges between the interval $\theta_{i}^{e}\in[0, 1]$.  
Under these assumptions, the electronic temperature evolution can be modeled using the following master equation 
\begin{equation} \label{eq:1}
	\frac{\partial T_{i}^{e}}{\partial t}=T_{\max }^{e} \sum_{\substack{j=1 \\ j \neq i}}^{N_{n}} K_{i j} \bigg\{\theta_{j}^{e}\left(1-\theta_{i}^{e}\right) \Gamma_{j \rightarrow i}  -\theta_{i}^{e}\left(1-\theta_{j}^{e}\right) \Gamma_{i \rightarrow j}  \bigg\}-\frac{G}{C_{e}}\left(T_{i}^{e}-T_{i}^{\mathrm{lat}}\right) + \frac{q_{i}}{C_{e}}.
\end{equation}

The master equation can be divided into three parts. 
The first term represents the temperature exchange due to the temperature gradient between neighboring atomic sites ($N_n$) contained between a sphere of radius $r_c = 2b$, with $b$ as the Burgers vector. 
The probability of energy exchange between two neighboring sites is represented by $\Gamma_{i \rightarrow j} = \exp(\theta_j - \theta_i)$.
The pair-wise thermal exchange coefficient $K_{i j}$ represents the probability exchange rate between two neighboring sites and can be computed either from \emph{ab-initio} calculations \cite{orhan2023electronic} or through experimentally determined thermodynamics properties \cite{ullah2019new,ponga2018unified}, given by
\begin{equation} \label{eq:2}
	K_{i j}=\frac{2 \kappa_{e} d}{C_{e} Z b^{2}},
\end{equation}
where $d=3$ is the dimension of the problem, and $Z$ is the coordination number in the reference (undeformed) configuration.
In our implementation, the temperature dependence of electronic thermal conductivity, $\kappa_{e}$, and heat capacity, $C_{e}$, are given by $\kappa_{e} = \frac{\kappa_{o} T_{i}^{e}}{T_{i}^{\mathrm{lat}}}$ and $C_{i} = \lambda T_{i}^{e}$, $\kappa_{o}$ is the thermal conductivity at room temperature.

The second term in the r.h.s of Eq. \ref{eq:1} represents the energy exchange between the electrons and phonons. 
The strength of this coupling is given by $G$ and is directly proportional to the temperature difference between the electronic ($T_i^e$) and lattice ($T_{i}^{\mathrm{lat}}$) subsystems. 

Finally, to model light-matter interaction, the third term in Eq. \ref{eq:1} is used to simulate the energy flux between the laser and surface plasmons, given by $q_i$.
The laser energy deposition, $q_{i}$, is described with a Gaussian temporal profile with the following expression \cite{chen2002modeling, shen2015effect}
\begin{equation} \label{eq:3}
	q_{i}=\sqrt{\frac{\beta}{\pi}} \frac{(1-R) F_{\mathrm{o}}}{t_{\mathrm{p}} L_{\mathrm{p}}}  \exp \left[-\left(\frac{x}{L_{\mathrm{p}}}\right)-\beta\left(\frac{t-2 t_{\mathrm{p}}}{t_{\mathrm{p}}}\right)^{2}\right]
\end{equation}
where $x$ is the scaled axial position of atom $i$, $F_{\mathrm{o}}$ is the peak laser fluence, $R$ is the reflectance, $t_{\mathrm{p}}$ is the pulse duration defined as the full width at half maximum, $L_p$ is the scaled optical penetration depth, and $\beta = {\mathrm{4ln(2)}}$.
In Eq. \ref{eq:3}, the laser pulse is applied on both the front and rear surface ($x = 0$ and $x = L$) at $t = {\mathrm{0}}$, and the peak laser fluence occurs at $t$ = 2$t_{\mathrm{p}}$.
The laser heat source attenuates away with an increase in depth by $\exp(-x/L_{p}$). 
It should be noted that both $x$ and $L_{\mathrm{p}}$ are scaled relative to the global maximum and minimum of the axial positions in the system. 
The scaled position $x$ is defined as $x = \frac{x_{\text{unscaled}} - x_{\text{min}}}{x_{\text{max}} - x_{\text{min}}}$, where $x_{\text{unscaled}}$ is the original axial position. 
Similarly, the scaled optical penetration depth $L_{\mathrm{p}}$ is given by $L_{\mathrm{p}} = \frac{L_{\mathrm{p, \text{unscaled}}}}{x_{\text{max}} - x_{\text{min}}}$.
Furthermore, the local lattice temperature is calculated as 
\begin{equation} \label{eq:4}
	T_{i}^{\mathrm{lat}}={\frac{2}{3 k_{B} N_{j}}} \sum_{j=1}^{N_{j}} {\frac{1}{2} \mathbf{m}_{j} \mathbf{v}_{j}\cdot \mathbf{v}_{j}},
\end{equation}
where $N_{j}$ represents the number of atoms defined by a cut-off radius $r_c$ around the $i-$th atom, $k_{B}$ is the Boltzmann constant, and $\mathbf{m}_{j}$ and $\mathbf{v}_{j}$ are the mass and velocity vector of the atom $j$, respectively.
To balance the energy exchange between electrons and phonons, the dynamics of the lattice is modified using a positive damping force, i.e.,
\begin{equation} \label{eq:5}
	m_{i} \dot{\mathbf{v}}_{i}=\mathbf{F}_{i}+\xi_{i} m_{i} \mathbf{v}_{i},
\end{equation}
where the strength of the damping force is computed at each time step for each atom as
\begin{equation} \label{eq:6}
	\xi_{i}=\frac{G V_{\text {atom }}\left(T_{i}^{e}-T_{i}^{\text {lat }}\right)}{m_{i} \mathbf{v}_{i}^2}.
\end{equation}
In Eq. \ref{eq:5}, $\mathbf{F}_{i}$ is the force acting on atom $i$ due to interatomic interaction, and the additional term was added to account for the electron-phonon energy coupling.
The material constants used for describing the electron subsystem in the $\ell2$T-MD model are $\lambda = 1065 \mathrm{~J} \cdot \mathrm{m}^{-3} \cdot\mathrm{K}^{-2}, \quad \kappa_{0}$ $=91 \mathrm{~W} \cdot \mathrm{m}^{-1} \cdot \mathrm{K}^{-1}, \quad G=3.6 \times 10^{17}$ $\mathrm{W} \cdot \mathrm{m}^{-3} \cdot\mathrm{K}^{-1}$, and $L_{p} = 13.5 \mathrm{nm}$ \cite{hohlfeld2000electron}.

\subsubsection{MD computational details}
\begin{figure}[h]
	\centering
	\includegraphics[width=1.0\textwidth]{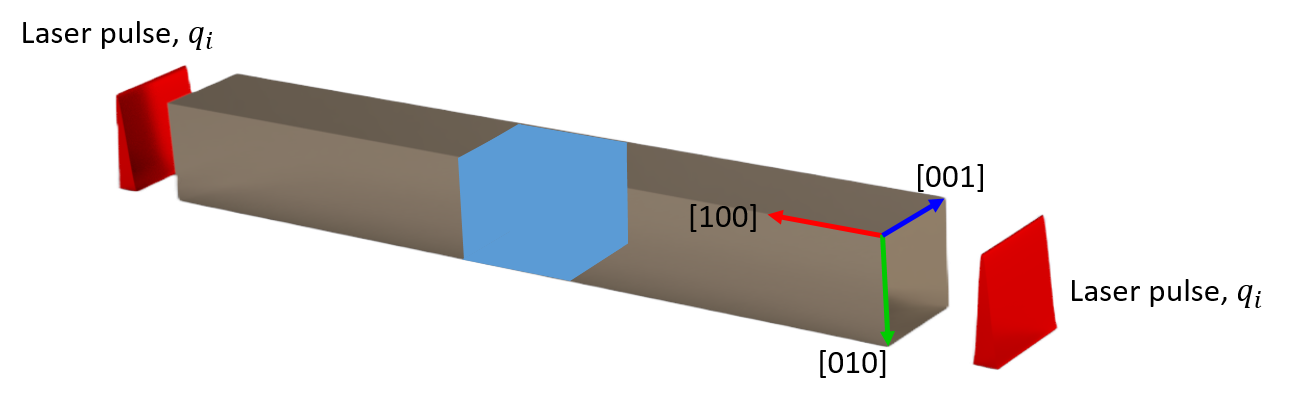}
	\caption{MD simulation schematic of the Ni sample irradiated on both ends along the [100] direction, aligned with the \textit{x}-axis, where the blue atoms represent the region selected for the dislocation analysis.}
	\label{fig:Schematic_MD_simulation}
\end{figure}
All $\ell2$T-MD model calculations were performed with the Large-scale Atomic Molecular Massively Parallel Simulator (LAMMPS) package \cite{plimpton1995fast, ullah2019new}.
The embedded atom method (EAM) \cite{daw1984embedded} has been widely employed to model metallic systems, and here, the DWY Ni EAM \cite{du2012construction} was used to model the atomic interaction due to good comparison with experimental data for shock response at elevated temperatures \cite{choi2018molecular}.
Single perfect Ni crystals were created with the \hkl[100], \hkl[010], and \hkl[001] crystallographic directions of the face-centered cubic phase aligned with the $x$, $y$, and $z$ axes of the simulation cell, as shown in Figure \ref{fig:Schematic_MD_simulation}.
We generated simulation cells with sizes $200 \times 5 \times 5$ nm$^3$ (containing over $0.5\times10^6$ atoms) to study the shock and spall behavior.
Furthermore, a separate set of MD simulations was conducted. The simulation domain was expanded to $200 \times 25 \times 25$ nm$^3$ (containing over $10\times10^6$ atoms) to study the failure mechanisms, including dislocation emission, void nucleation, growth, and coalescence. 
In one of these samples, an initial nanovoid with radius $r_\mathrm{v} = 2$ nm was included to study the dependency of the spall strength with porosity. 

A traction-free boundary condition was applied in the direction parallel to the laser pulse ($x$), while periodic boundary conditions were applied to the transverse ($y$ and $z$) directions.
To obtain reproducible results, the initial structures were equilibrated at 300 K and zero pressure for 100 ps based on an isothermal-isobaric ensemble (NPT) \cite{martyna1994constant} to achieve a thermodynamic equilibrium state before the dynamic loading.
The time-step in MD simulations was chosen to be $\Delta t = 0.1$ fs to ensure numerical stability.
The Ovito software was used to visualize the atomic trajectory and analyze the generation of voids \cite{stukowski2009visualization}.
To investigate the pressure and temperature over the sample, we created several slices along the $x-$direction using the $1D$ binning analysis, with a width size of 1 nm.
Physical quantities, such as velocity, temperature, density, and stress fields, were obtained by averaging the atoms within each slice for a short period of time. 
Local deformation and structure were determined using the common neighbor analysis (CNA) \cite{honeycutt1987molecular}, dislocation extraction algorithm (DXA) \cite{stukowski2010extracting}, and by calculating the coordination numbers \cite{stukowski2009visualization}.
\subsection{Analytical calculation of thermoelastic stress}
The rapid heating of a material by a laser pulse results in the propagation of thermoelastic waves due to the material's thermal expansion near the heated surfaces. 
Since the time scale associated with laser pulses is very short, these waves are characterized by a very high strain rate coupled with temperature distribution.
This coupling effect leads to stress attenuation and localized temperature variation \cite{wang2001thermoelastic}.
However, using the classical Fourier law is not valid as these stress and thermal waves propagate at a finite speed \cite{tzou2014macro}.
One approach to address this issue is to modify the classical heat conduction equation by incorporating a thermal relaxation time \cite{wang2001thermoelastic}. 
Alternatively, electronic effects can be incorporated to account for the absorption of laser energy by electrons and subsequent heat conduction to the lattice \cite{wang2002thermoelastic}.

The thermo-elastic stress can be described analytically by considering the laser pulse energy as a Fourier series \cite{wang2002thermoelastic}, and calculating the electronic and lattice temperature using the two-temperature model (TTM) \cite{kagnov1957relaxation,anisimov1974electron}, and thermal expansion is considered through the equation of motion. 
For a one-dimensional system with a linear elastic and isotropic behavior, the governing equations for electronic and lattice temperature, $T^{e}$ and $T^{\mathrm{lat}}$, respectively, and the lattice displacement, $u$, consist of three coupled partial differential equations \cite{lord1967generalized, brorson1987femtosecond}.
\begin{equation} \label{eq:13}
	C_{e} \frac{\partial T^{e}}{\partial t} = \frac{\partial}{\partial x}(\kappa_{e} \frac{\partial T^{e}}{\partial x})-G(T^{e}-T^{\mathrm{lat}})+\beta I(t) e^{-\beta x}
\end{equation}
\begin{equation} \label{eq:14}
	C_{\mathrm{lat}} \frac{\partial T^{\mathrm{lat}}}{\partial t} = G(T^{e}-T^{\mathrm{lat}})-K \beta_{T} T_0 \frac{\partial^2 u}{\partial x \partial t}
\end{equation}
\begin{equation} \label{eq:15}
	\rho \frac{\partial^2 u}{\partial t^2} = (K + \frac{4 \mu}{3}) \frac{\partial^2 u}{\partial x^2}-K \beta_{T} \frac{\partial T^{\mathrm{lat}}}{\partial x}
\end{equation}
where $x$ is the coordinate orthogonal to the heated surface, $C_{e}$ and $C_{\mathrm{lat}}$ are the volumetric electronic and lattice heat capacities, respectively, $\kappa_e$ is the heat conductivity, $G$ is the electron-phonon coupling strength, $\beta$ is the optical absorption coefficient, $I(t)$ is the temporal laser pulse profile, $K$ and $\mu$ are the bulk and shear moduli, $\rho$ is the density, $\beta_{T}$  is the volumetric thermal expansion coefficient, and $T_{0}$ is the initial temperature of the material.

Eq. \ref{eq:13} describes heat diffusion, electron-phonon energy exchange, and energy absorbed by the electrons from the laser energy. 
No diffusion term is considered in Eq. \ref{eq:14} since heat diffuses much more rapidly through electrons than through the lattice \cite{brorson1987femtosecond}. 
Lastly, Eq. \ref{eq:15} represents the equilibrium equation in terms of lattice displacement, including thermal expansion effects.
Since the elastic wave is generated and propagated by thermal expansion and lattice vibration, only the lattice temperature appears in this equation.

\begin{figure}[h!]
	\centering
	\begin{subfigure}{0.46\textwidth} \centering
		\includegraphics[width=0.99\textwidth]{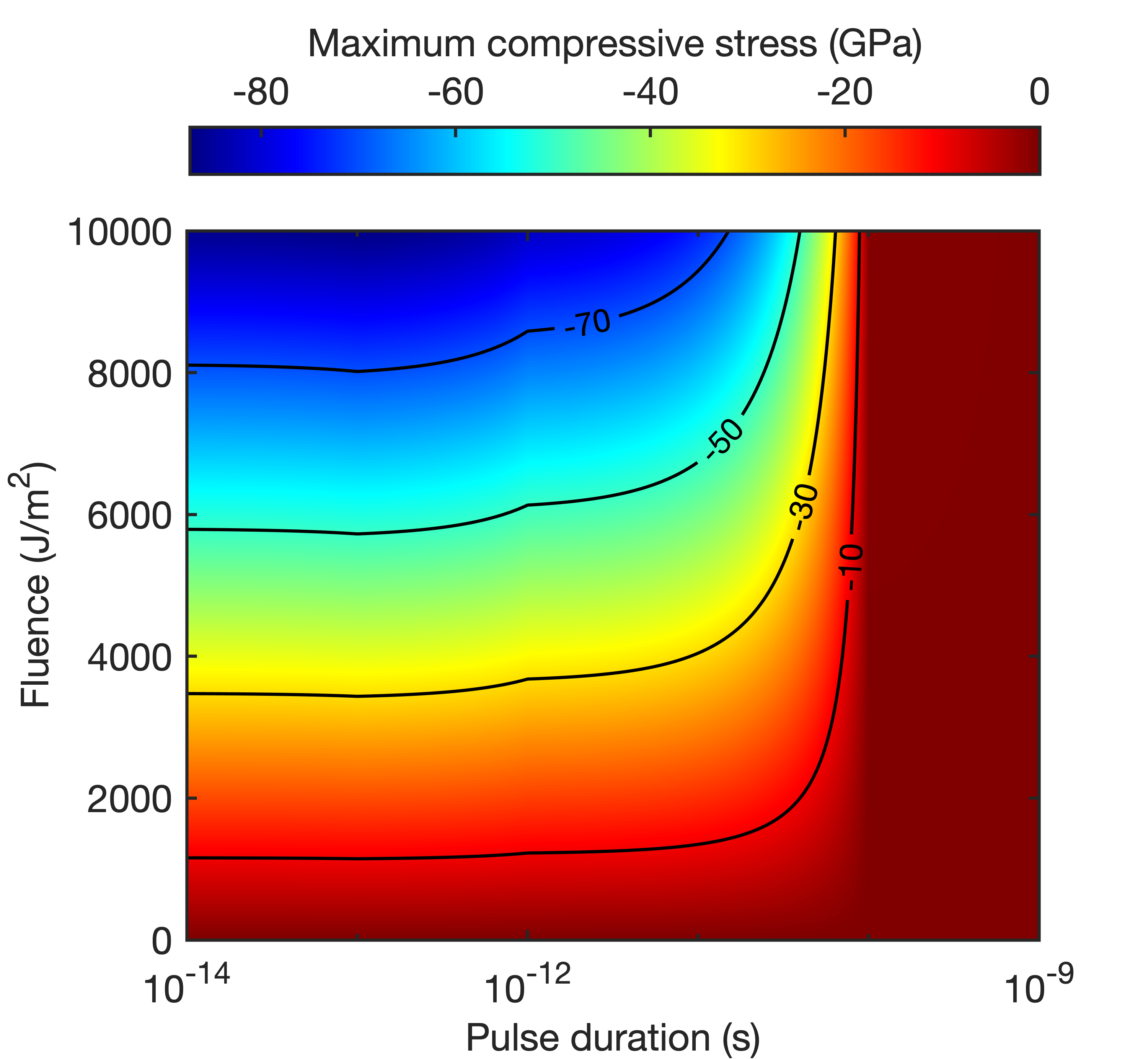}
		\caption{}
	\end{subfigure}
	\begin{subfigure}{0.46\textwidth} \centering
		\centering
		\includegraphics[width=0.99\textwidth]{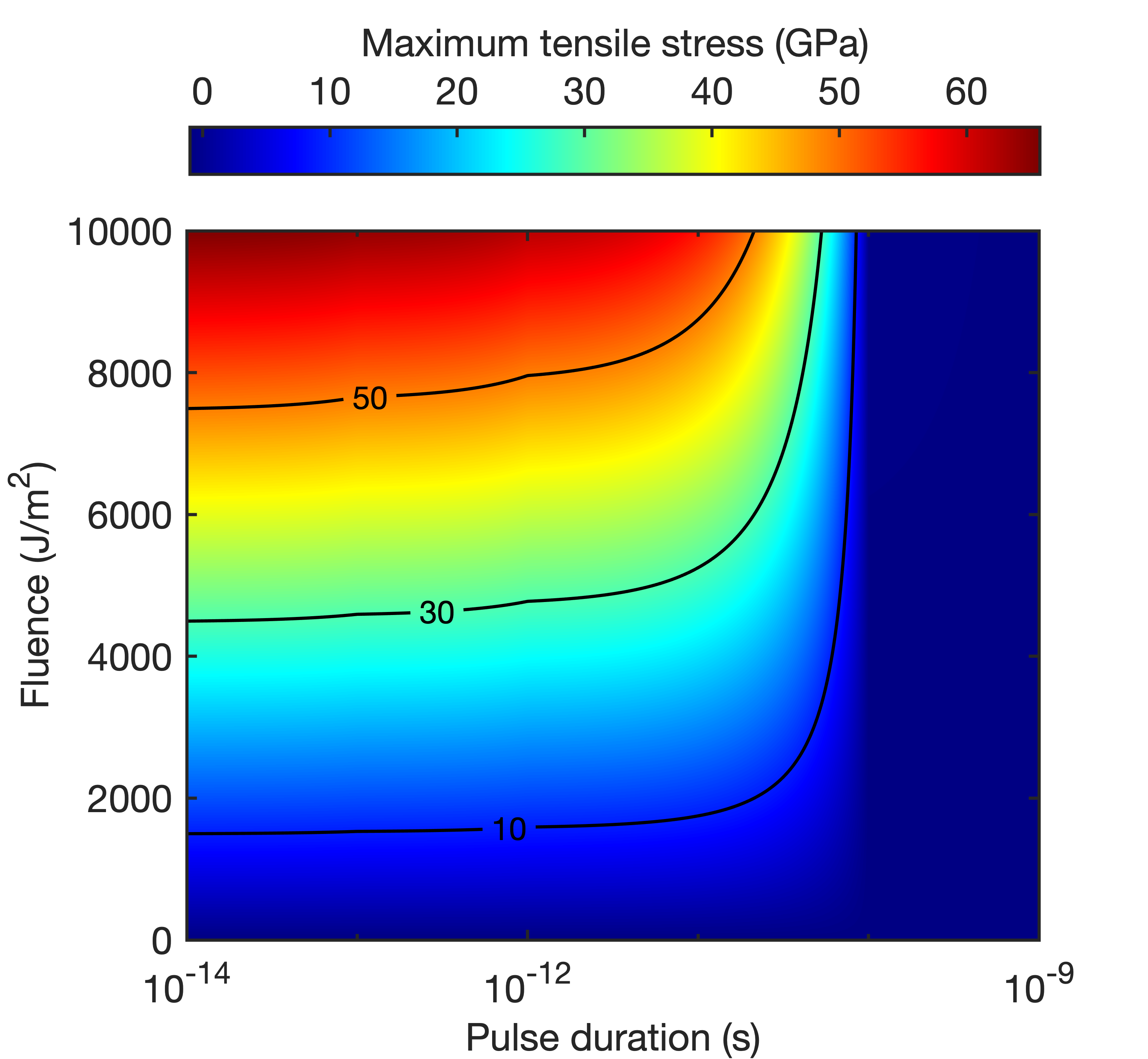}
		\caption{}
	\end{subfigure}
	\caption{Analytical calculation of the maximum (a) compressive and (b) tensile stress as a function of laser fluence and pulse duration.}
	\label{fig:analytical_calculation_comp_ten}
\end{figure}

Initially, the sample is assumed to have a uniform temperature, no displacement, and no stress before heating, with the first-order time derivatives of temperature and displacement set to zero. 
The sample's surface is presumed to be thermally insulated and stress-free.
When $x$ approaches infinity, the sample is fixed without an increase in temperature or stress.
The analytical solution to the coupled equations is obtained by expressing the laser pulse energy as a Fourier series
\begin{equation} \label{eq:16}
	I(t)=a_0+\sum_{i=1}^{\infty}\left(a_i \cos (\omega t)+b_i \sin (\omega t)\right)
\end{equation}
where $a_0$, $a_i$ and $b_i$ are coefficients of Fourier series, $\omega$ = $i \omega_0$ and $\omega_0$ = $2 \pi f_0$ where $f_0$ is the laser pulse repetition rate, and assumed to be large enough such that the solution of the temperature and stress wave corresponds to that of a single laser pulse. 
The temperature increase induced by $I(t)$ consists of two components: a steady temperature increase caused by $a_0$, and a temporal part resulting from the other terms in Eq. \ref{eq:16}.
Due to the linear relationship between $I(t)$ and the lattice displacement, the thermoelastic stress can be expressed as
\begin{equation} \label{eq:17}
	\sigma(x,t)=\sum_{i=1}^{\infty}\left(a_i \operatorname{Re}\left(\tilde{\sigma}_i\right)+b_i \operatorname{Im}\left(\tilde{\sigma}_i\right)\right),
\end{equation}
with $\tilde{\sigma}_i$ representing the stress induced by the laser pulse with a complex intensity of $e^{i \omega t}$.
$\operatorname{Re}$ and $\operatorname{Im}$ are the real and imaginary components of a complex number.
Similarly, by expressing the lattice displacement in terms of $e^{i \omega t}$, a general homogeneous solution of the form $A_i$ $e^{k_j x}$ can be obtained, with $A_i$ and $k_j$ solved using appropriate boundary conditions. 
As a result, the analytical solution for thermoelastic stress is given by
\begin{equation} \label{eq:18}
	\begin{split}
		\sigma_i&=(K + \frac{4 \mu}{3})\cdot(e^{k_{1,i}x}(B_{1,i} k_{1,i}+v A_{1,i})+e^{k_{2,i}x}(B_{2,i} k_{2,i}+v A_{2,i})) \\
		& +e^{-\beta x}(-\beta B_{\mathrm{p},i}+v A_{\mathrm{p},i})),
	\end{split}
\end{equation}
where $v = {-K \beta_T}/{(K + \frac{4 \mu}{3})}$ and the parameters $B_{1,i}$, $B_{2,i}$, $A_{1,i}$, $A_{2,i}$, $k_{1,i}$, and $k_{2,i}$ are solved using the boundary conditions.
The particular solutions, $A_{\mathrm{p}}$ and $B_{\mathrm{p}}$, are obtained from Eqs. \ref{eq:13}-\ref{eq:15}.
The solutions to the Fourier parameters are provided in Appendix A.
The final solution is obtained by substituting $\tilde{\sigma}_i = \sigma_i e^{i \omega t}$ into Eq. \ref{eq:17}.
The material properties used to analytical calculate the thermoelastic stress in nickel are given in Table \ref{tbl:2}. 
Analytically solving the TTM provides an insight into the relationship between laser fluence and pulse duration with stress state, as shown in Figure \ref{fig:analytical_calculation_comp_ten}.

Here, we are assuming a single laser pulse, but in principle, by using the proposed two-pulse laser approach we could double the peak stress due to the interaction of stress waves.
Hence, in theory, spallation can be induced at a lower fluence in compared to the conventional single laser approach. 
As noted in the literature \cite{shugaev2021laser}, increasing pulse duration and laser fluence can lead to thermal processes such as melting, phase explosion, and boiling.
Therefore, compared to the conventional method, the proposed method may avoid these processes by requiring less laser energy to produce higher hydrostatic pressure.
\section{Results} 
\subsection{Experimental exploration of two-pulse laser spall approach}
\begin{figure}[h!]
	\centering
	\begin{subfigure}{0.46\textwidth} \centering
		\includegraphics[width=0.99\textwidth]{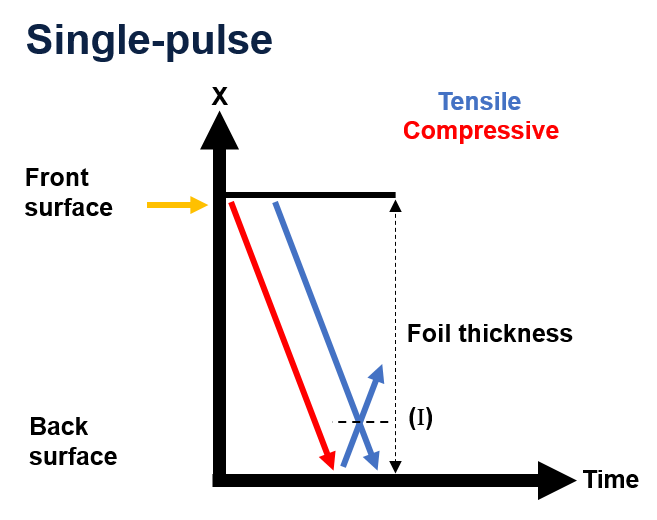}
		\caption{}
		\label{fig:Idealized_shock_wave_behaviour_single_pulse_laser_approach}
	\end{subfigure}
	\begin{subfigure}{0.46\textwidth} \centering
		\centering
		\includegraphics[width=0.99\textwidth]{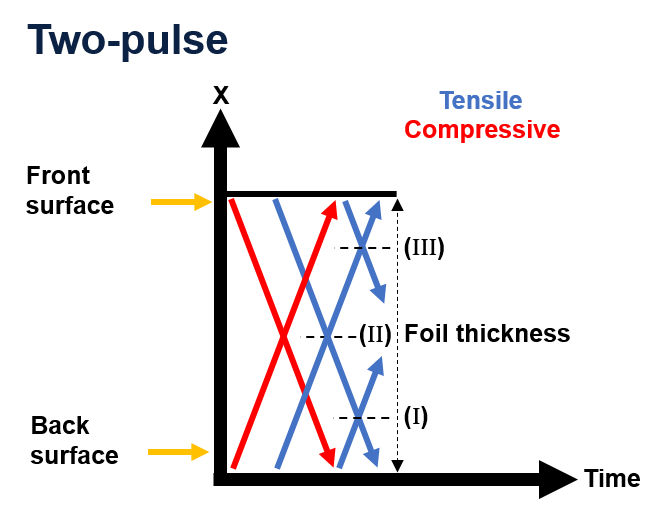}
		\caption{}
		\label{fig:Idealized_shock_wave_behaviour_two_pulse_laser_approach}
	\end{subfigure}
	\caption{Idealized shockwave behavior under (a) single- and (b) two-pulse laser spallation approach. These position-time plots show the increased number of potential spall location with the two-pulse approach arising from the interaction of the unloading tensile waves with each other, (II), or the interaction of the unloading tensile wave with the reflected compressive wave, (I) and (III).}
	\label{fig:Idealized_shock_wave_behaviour_single_two_pulse_laser_approach}
\end{figure}

Before presenting the experimental results, it is essential to first understand the idealized shockwave behavior associated with both the single- and two-pulse laser approaches, as illustrated in Figure \ref{fig:Idealized_shock_wave_behaviour_single_two_pulse_laser_approach}.
The shockwave behavior under the single-pulse approach has been extensively studied in the literature \cite{eliezer1990laser, shugaev2019thermodynamic, shugaev2021laser}.
It is well established that spall failure occurs due to the generation of large tensile hydrostatic stress, which arises from the interaction between the reflected compressive stress wave and the incoming rarefaction (or unloading tensile) waves, represented here for simplicity by a single unloading wave.
This interaction typically occurs near the unloaded surface.
It should be noted that this behavior is typically observed under lower laser fluence conditions.
At extremely high laser fluences, the unloading tensile component may vanish due to energy absorption in the plasma state \cite{romain2002pressure}. %
When a secondary laser pulse is introduced in the two-pulse approach, an additional set of compressive and unloading tensile waves is created. 
The interaction of these waves produces a state of hydrostatic stress, potentially leading to spall at three distinct locations: near the front surface, the back surface, and/or the center of the specimen.
With this idealized shockwave behavior now clarified, we will proceed to investigate the two-pulse laser spallation method by comparing its observable features with those of the single-pulse laser spallation approach.
\subsubsection{Spall observation : Microscale-thick Ni foil}
\begin{figure}[h!]
	\centering
	\includegraphics[width=1.0\textwidth]{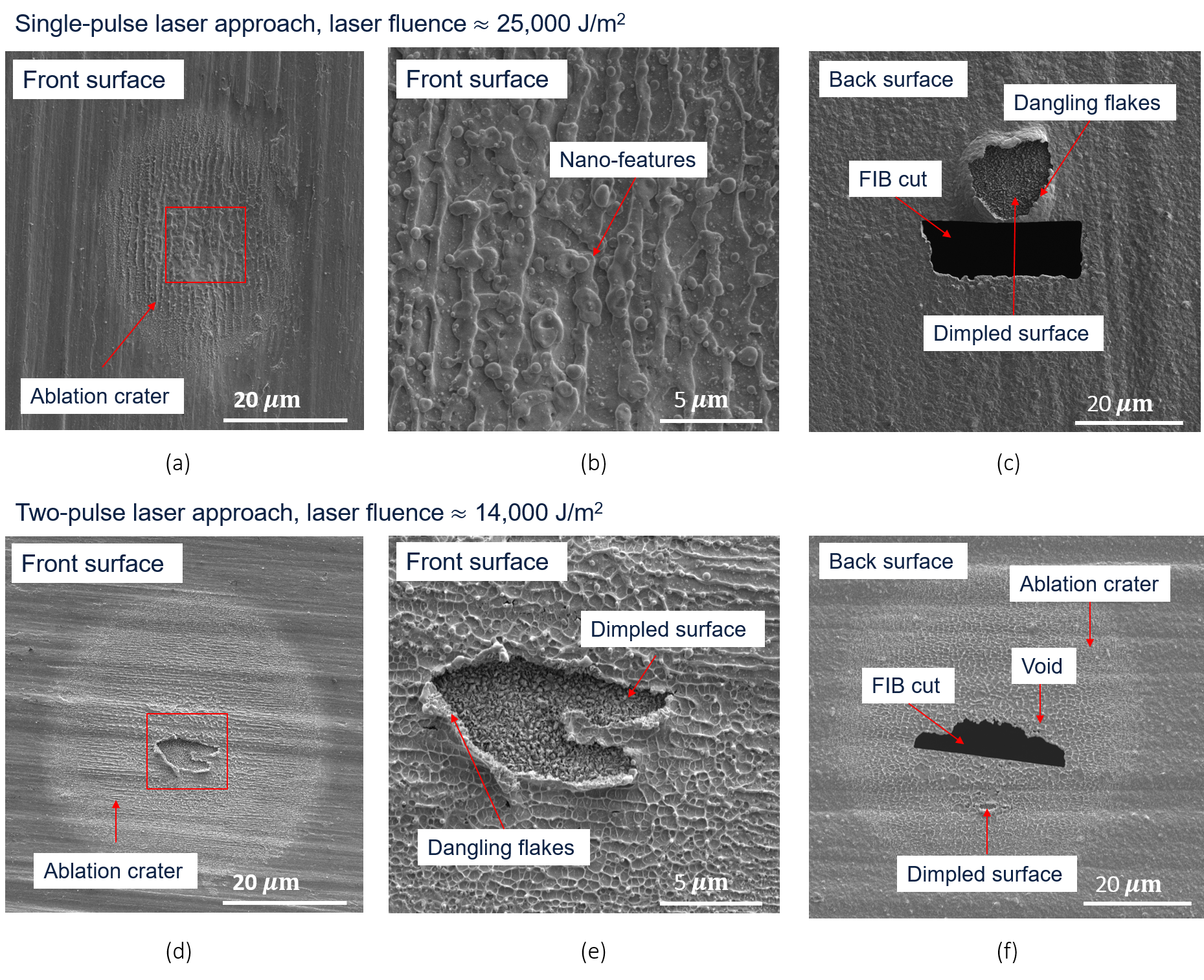}
	\caption{SEM images of the surface features produced as a result of the single- and two-pulse laser spall approach on the microscale-thick Ni foil with (a,d) displaying the front surface, (b,e) a zoomed-in view of the front surface, and (c,f) the back surface. These images show classical ductile failure mechanisms characterized by dimpling and flaking on the specimen's surface.}
	\label{fig:Surface_observation_single_two_pulse_laser_approach}
\end{figure}
\begin{figure}[h!]
	\centering
	\includegraphics[width=1.0\textwidth]{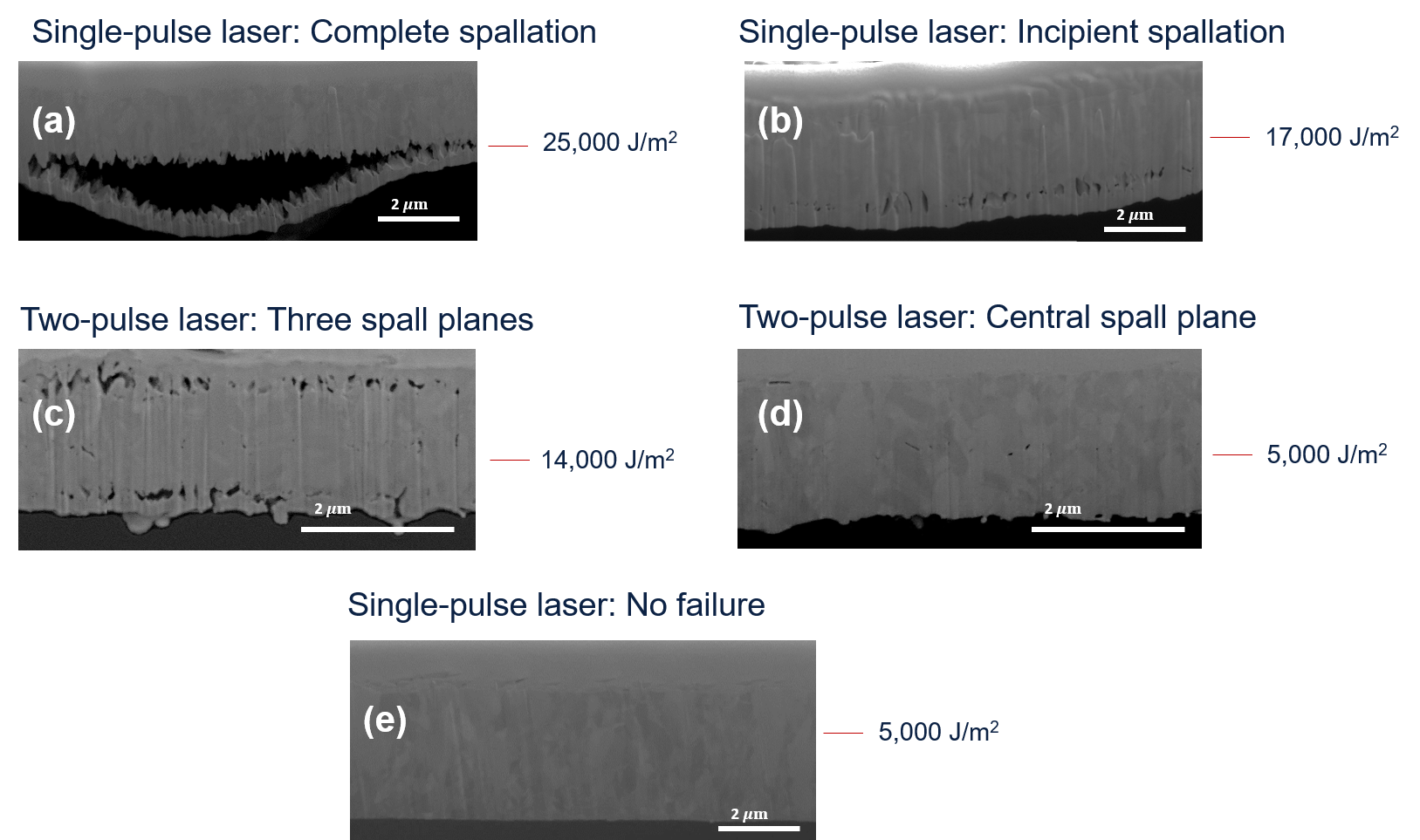}
	\caption{SEM images of the exposed sub-surface milled by FIB showing the internal damage in the microscale-thick Ni foil illuminated by the single- and two-pulse laser approach across different laser fluences. At the lowest laser fluence, central voids are observed in the two-pulse approach, whereas no failure is evident in the single-pulse approach under similar conditions.}
	\label{fig:Microscale_thick_Ni_spall_failure_observation_single_two_pulse_laser_approach}
\end{figure}

Figure \ref{fig:Surface_observation_single_two_pulse_laser_approach} illustrates both the surface observation of the microscale-thick Ni foil under the single and two-pulse configurations.
Once irradiated at a laser fluence above the threshold for spall failure, the rapid increase in surface temperature leads to the formation of an intricate network of nano-spikes and nano-pits within the ablation crater (cf., Figure \ref{fig:Microscale_thick_Ni_spall_failure_observation_single_two_pulse_laser_approach}(a)-(b) $\&$ (d)-(f)).
The formation of these nanostructures has been attributed to re-solidified liquid Ni that melted initially \cite{Ionin2015,Amoruso2007,Temnov2020}.
The complex patterns arise from the intense jetting experienced due to extreme hydrodynamic expansion of the supercritical Ni liquid \cite{Ionin2015}.
In the conventional single-pulse laser approach, complete spall failure results in the exposure of the spall plane on the unloaded surface as shown in Figure \ref{fig:Surface_observation_single_two_pulse_laser_approach}(c), and highlighted in the literature \cite{eliezer1990laser,Jarmakani2010,righi2021towards,Zhu2024}.
Similarly, in the proposed approach, the same phenomena take place on both the front and back surfaces, as shown in Figure \ref{fig:Surface_observation_single_two_pulse_laser_approach}(d)-(f).
The exposed fracture surface, i.e., spall plane, shows a dimpled morphology as a result of the nucleation, growth, and coalescence of voids \cite{meyers1983dynamic}.
Interestingly, dangling flakes can be observed on the exposed fracture surface in both approaches, as shown in Figure \ref{fig:Surface_observation_single_two_pulse_laser_approach}.
Flaking, another form of ductile failure, occurs due to the separation along grain boundaries \cite{Jarmakani2010,righi2021towards}.
These features are all typical in laser-metal interaction due to the complex interplay between the thermal and mechanical induced effects.
Thus, both approaches result in similar surface characteristics; however, the key difference is that the two-pulse laser approach produces spall features, i.e., ablation crater, flaking, and exposed spall planes, on both the front and back surfaces.

In order to further examine spall behaviour under the single- and two-pulse laser approaches, the sub-surface of the illuminated microscale-thick Ni foil was examined using a focused ion beam (FIB) to reveal any internal damage as shown in Figure \ref{fig:Microscale_thick_Ni_spall_failure_observation_single_two_pulse_laser_approach}.
Under the single-pulse laser approach, at sufficiently large laser fluences (= 25,000 $\mathrm{J \cdot m^{-2}}$, see Figure \ref{fig:Microscale_thick_Ni_spall_failure_observation_single_two_pulse_laser_approach}(a)), complete spallation takes place where nucleation, growth, and coalescence of voids lead to the formation of a new surface.
Incipient spall, the precursor to complete spall failure, was achieved under the single-pulse laser approach by reducing the laser fluence (= 17,000 $\mathrm{J \cdot m^{-2}}$, see Figure \ref{fig:Microscale_thick_Ni_spall_failure_observation_single_two_pulse_laser_approach}(b)) where a series of voids can be seen close to the back surface, i.e., at location (I) in Figure \ref{fig:Idealized_shock_wave_behaviour_single_pulse_laser_approach}.

Once the laser fluence for incipient spallation was established, we turned our attention to the two-pulse laser approach (= $14,000~\mathrm{J \cdot m^{-2}}$, see Figure \ref{fig:Microscale_thick_Ni_spall_failure_observation_single_two_pulse_laser_approach}(c)) where three distinct spall failure sites were observed: (i) near the top surface, (ii) near the bottom surface, and (iii) at the center of the specimen.
The interaction of the unloading tensile waves generated sufficient pressure to nucleate voids at the specimen's center.
After reaching the center, these unloading waves propagate towards the free surfaces, where they interact with the reflected compressive waves, leading to failure near both the top and bottom surfaces.
Notably, voids near the surfaces are larger than those at the center.
This difference can be attributed to the hydrostatic tensile stress generated by the interaction of the unloading tensile waves with the reflected compressive waves being greater than the stress produced by the superposition of the unloading tensile waves alone, i.e., the hydrostatic stress developed at locations (I) and (III) is greater than at (II) in Figure \ref{fig:Idealized_shock_wave_behaviour_two_pulse_laser_approach}.

Figures \ref{fig:Microscale_thick_Ni_spall_failure_observation_single_two_pulse_laser_approach}(d)--(e) show that at sufficiently low laser fluences (= $5,000 ~\mathrm{J \cdot m^{-2}}$), the two-pulse laser approach induces failure at the center of the specimen due to the interaction of unloading tensile waves, whereas the single-pulse approach does not result in any failure. %
A comparison of both approaches reveals that the hydrostatic stress generated near the surfaces was insufficient to cause failure, as no spallation was observed in those regions.
This phenomenon is likely due to a combination of factors: (a) the reduced magnitude of the compressive waves as laser fluence decreases, as demonstrated in the analytical solution shown in Figure \ref{fig:analytical_calculation_comp_ten}, (b) the attenuation of the unloading tensile waves as they propagate from the center towards the surfaces \cite{CuqLelandais2009,hu2009high,Zhang2022}, and (c) the onset of void nucleation at the center, which reduces the magnitude of the tensile stresses \cite{leveugle2004photomechanical}. 

These exploratory experimental results demonstrate that the two-pulse laser spallation method can indeed induce failure at the center of the specimen, similar to the plate-impact method. 
However, this failure does not produce complete spallation at the center, as it only results in the nucleation and growth of voids.
This incomplete spall failure may be attributed to wave attenuation \cite{CuqLelandais2009, hu2009high, Zhang2022}, which reduces the stress generated by the interaction of the unloading tensile waves, making it insufficient to cause more pronounced failure at the center, i.e., the hydrostatic tensile stress generated at location (II) in Figure \ref{fig:Idealized_shock_wave_behaviour_two_pulse_laser_approach} is insufficient to produce complete spall failure.
\subsubsection{Spall observation : nanoscale-thick Ni foil}
\begin{figure}[h!]
	\centering
	\includegraphics[width=1.0\textwidth]{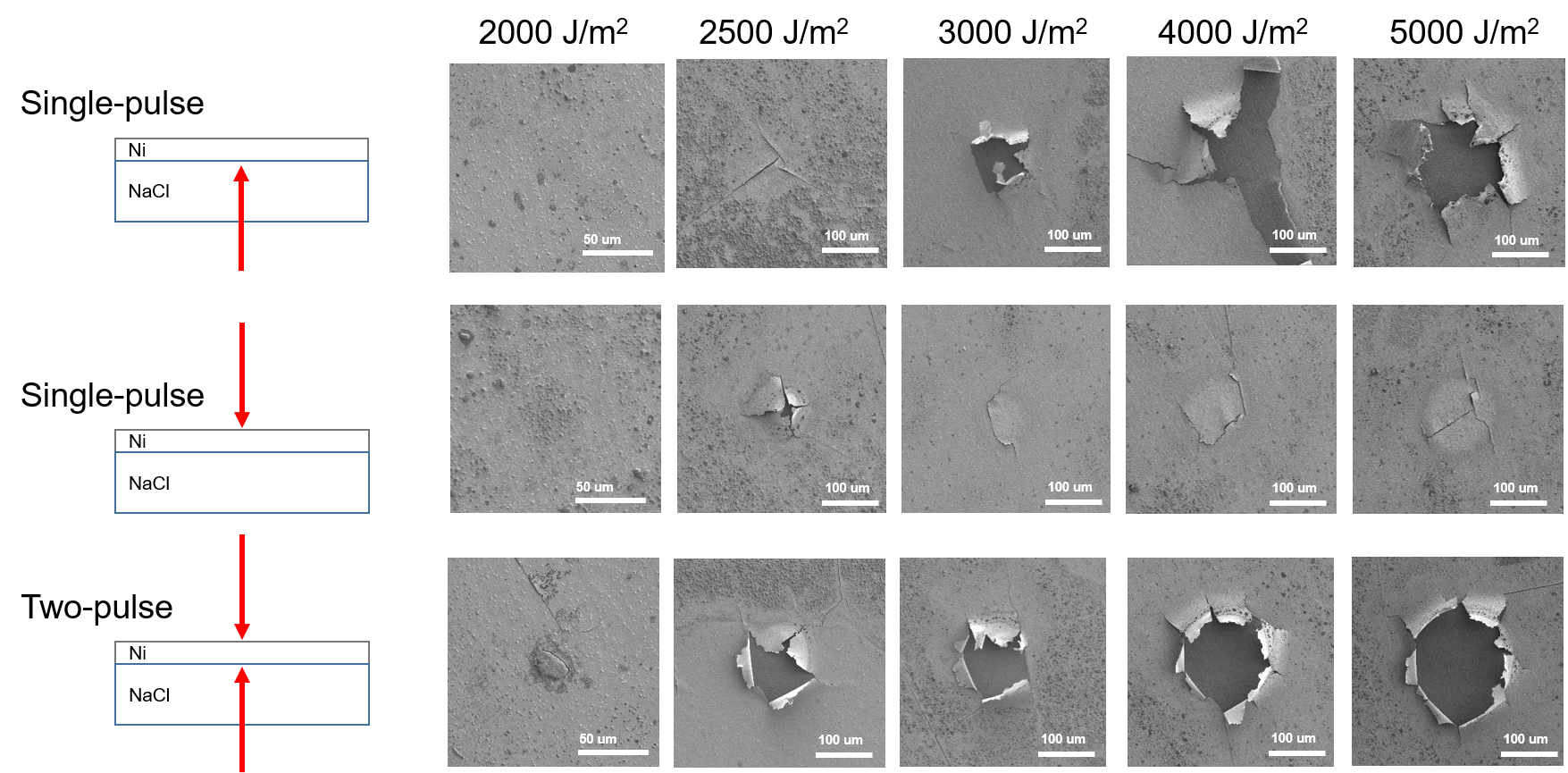}
	\caption{SEM images of the nanoscale-thick Ni foils subjected to the single- and two-pulse laser spallation approaches, showing the transition in spall behavior with varying laser fluence. In these images, the Ni surface (visible as a silver-toned area) is displayed, with the underlying NaCl substrate becoming exposed at higher laser fluences due to complete spallation.}
	\label{fig:Nanoscale_thick_Ni_spall_failure_observation_single_two_pulse_laser_approach}
\end{figure}
\begin{figure}[h!]
	\centering
	\includegraphics[width=1.0\textwidth]{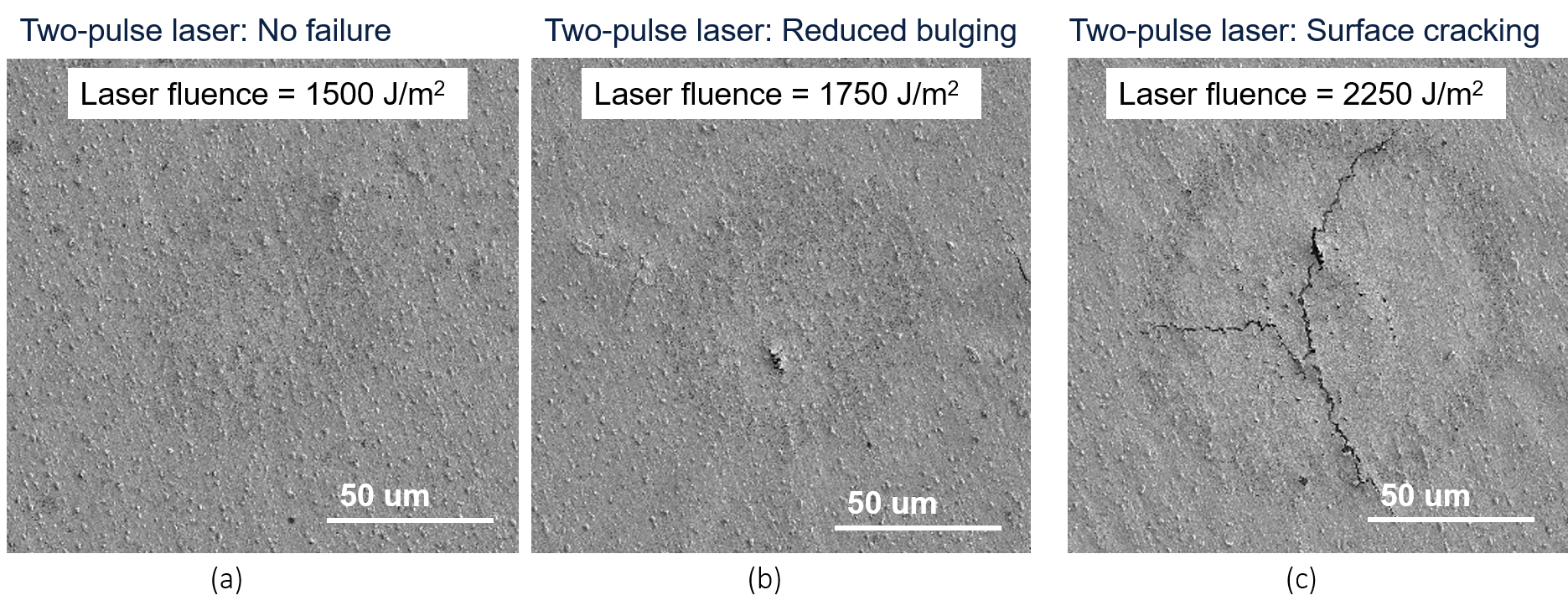}
	\caption{SEM images of the nanoscale-thick Ni foils subjected to the two-pulse laser spallation approach. Here, we see notably spall features at lower laser fluences in comparison to the single-pulse laser approach. In these images, the Ni surface (visible as a silver-toned area) is displayed, with the underlying NaCl substrate beneath the Ni foil.}
	\label{fig:Nanoscale_thick_Ni_spall_failure_observation_two_pulse_laser_approach_1500_1750_2250}
\end{figure}

Having established that the two-pulse laser spallation approach can induce failure at the center of the sample, away from the free surfaces, we turned our attention to exploring the potential effects of wave attenuation. 
To do this, we studied the behavior of nanoscale-thick Ni foils under both the single- and two-pulse approaches, as illustrated in Figure \ref{fig:Nanoscale_thick_Ni_spall_failure_observation_single_two_pulse_laser_approach}.
Comparing the results at 5,000 $\mathrm{J \cdot m^{-2}}$ from Figures \ref{fig:Microscale_thick_Ni_spall_failure_observation_single_two_pulse_laser_approach}(d)--(e) and \ref{fig:Nanoscale_thick_Ni_spall_failure_observation_single_two_pulse_laser_approach}, it is evident that the nanoscale-thick Ni foil exhibits more pronounced failure.
This is likely due to reduced wave attenuation in the nanoscale foil, leading to enhanced hydrostatic stress development.
In the single-pulse approach, where the nanoscale foil was irradiated both directly and through the glass substrate, differences in failure patterns were observed as the spall plane shifts between the two surfaces, as shown in Figure \ref{fig:Idealized_shock_wave_behaviour_single_two_pulse_laser_approach}(a).
As the laser fluence decreased, the degree of spall failure diminished from complete spall of the ablation crater to partial exposure of the illuminated surface and eventually to no observable failure.
The spall failure threshold for the single-pulse approach was identified at 2,500 $\mathrm{J \cdot m^{-2}}$.
The two-pulse laser behavior of the Ni nanofoil is shown in Fig.~\ref{fig:Nanoscale_thick_Ni_spall_failure_observation_two_pulse_laser_approach_1500_1750_2250}.
In contrast to the single pulse approach, the two-pulse laser approach achieved spall failure at lower fluences. 
Notable spall features include (a) surface cracking at 2,250 $\mathrm{J \cdot m^{-2}}$, (b) surface bulging at 2,000 $\mathrm{J \cdot m^{-2}}$, and (c) reduced bulging at 1,750 $\mathrm{J \cdot m^{-2}}$ - marking the spall threshold for the two-pulse laser approach. No appreciable failure was identified below 1,500 $\mathrm{J \cdot m^{-2}}$.

Considering the laser fluence thresholds for spall failure in the microscale- and nanoscale-thick Ni foils (cf. Figures \ref{fig:Microscale_thick_Ni_spall_failure_observation_single_two_pulse_laser_approach}(e) and \ref{fig:Nanoscale_thick_Ni_spall_failure_observation_single_two_pulse_laser_approach}), the wave attenuation factor can be roughly estimated as the ratio between the fluences, i.e., 35\%. 
This indicates that the fluence required for spallation in the nanoscale-thick Ni foil is 35\% of that required for the microscale-thick Ni foil.
Hence, these results show that instead of irradiating a single surface of a specimen with a larger fluence, we can divide this fluence into two pulses.
This approach reduces the thermal effects associated with higher fluence levels \cite{shugaev2021laser}.
At the same time, provided that wave attenuation is addressed, the interaction of the unloading tensile waves from the two-pulse configuration can generate sufficient hydrostatic stress to induce nanovoid growth and coalesence in micro-sized specimens and spall failure in nanofoils.
\subsubsection{Spall strength and strain rate calculation}
It has been shown in the literature that, under the single-pulse laser approach, spall strength can be obtained through the spall thickness, i.e., the distance between (I) and the back surface in Figure \ref{fig:Idealized_shock_wave_behaviour_single_pulse_laser_approach} \cite{gilath1988laser, Jarmakani2010}.
Assuming that the unloading tensile stress wave travels through the target at a constant velocity and has a perfectly triangular shape, the spall thickness $\Delta$, foil thickness $L$, pulse duration $t_p$, sound velocity $c_0$ and the shockwave velocity $U_s$ are related by the following expression \cite{gilath1988laser}:
\begin{equation} \label{eq:19}
	\Delta = \frac{U_s -  c_0}{U_s +  c_0}L + \frac{t_p U_s c_0}{U_s +  c_0},
\end{equation}
Given that $\Delta \approx$ 200 --300 nm (estimated from Figures \ref{fig:Microscale_thick_Ni_spall_failure_observation_single_two_pulse_laser_approach}(a)-(b)), $L=2.5 ~\mu$m, $t_p = 350$ fs, $c_0$ = 4,650 $\mathrm{m \cdot s^{-1}}$, the calculated $U_s$ is 5,680 $\mathrm{m \cdot s^{-1}}$.
Using the Hugoniot relationship, the calculated particle velocity $U_p = (U_s - c_o)/s$ = 710 $\mathrm{m \cdot s^{-1}}$ and shock pressure $P_o = \rho U_s U_p$ = 35.8 GPa. 
For a triangular pressure pulse, the spall strength and strain rate can be expressed as $\sigma_\mathrm{spall} = 2(P_o - \rho c_o U_p)$ and $\dot{e} = U_p/L$, respectively \cite{gilath1988laser}.
The spall strength calculated from the spall thickness in the microscale-thick Ni foil is $\sim12.9$ GPa with an estimated strain rate of $2.8\times 10^8 ~\mathrm{s}^{-1}$.
\subsection{Molecular dynamics analysis of shock \& spall behavior}
Next, we examine the interaction between Ni and two laser pulses with a fluence of 1,250 $\mathrm{J \cdot m^{-2}}$, which corresponds to the MD observed spallation threshold, with a pulse duration of $t_{p}$ = 0.1 ps and an initial temperature of 300 K.
It is important to note that the spall threshold identified for the nanoscale-thick Ni foil, 1750 $\mathrm{J \cdot m^{-2}}$, does not represent the actual absorbed laser fluence, as the reflectivity of Ni, approximately 77$\%$ according to the literature \cite{Johnson1974}, would need to be taken into account. 
Consequently, the actual experimental absorbed laser fluence is calculated to be 402.5 $\mathrm{J \cdot m^{-2}}$. 
As expected, this is significantly lower than the MD-observed threshold for spallation, reflecting the disparity between experimental and computational conditions.

\begin{figure}[h]
	\centering
	\includegraphics[width=1.0\textwidth]{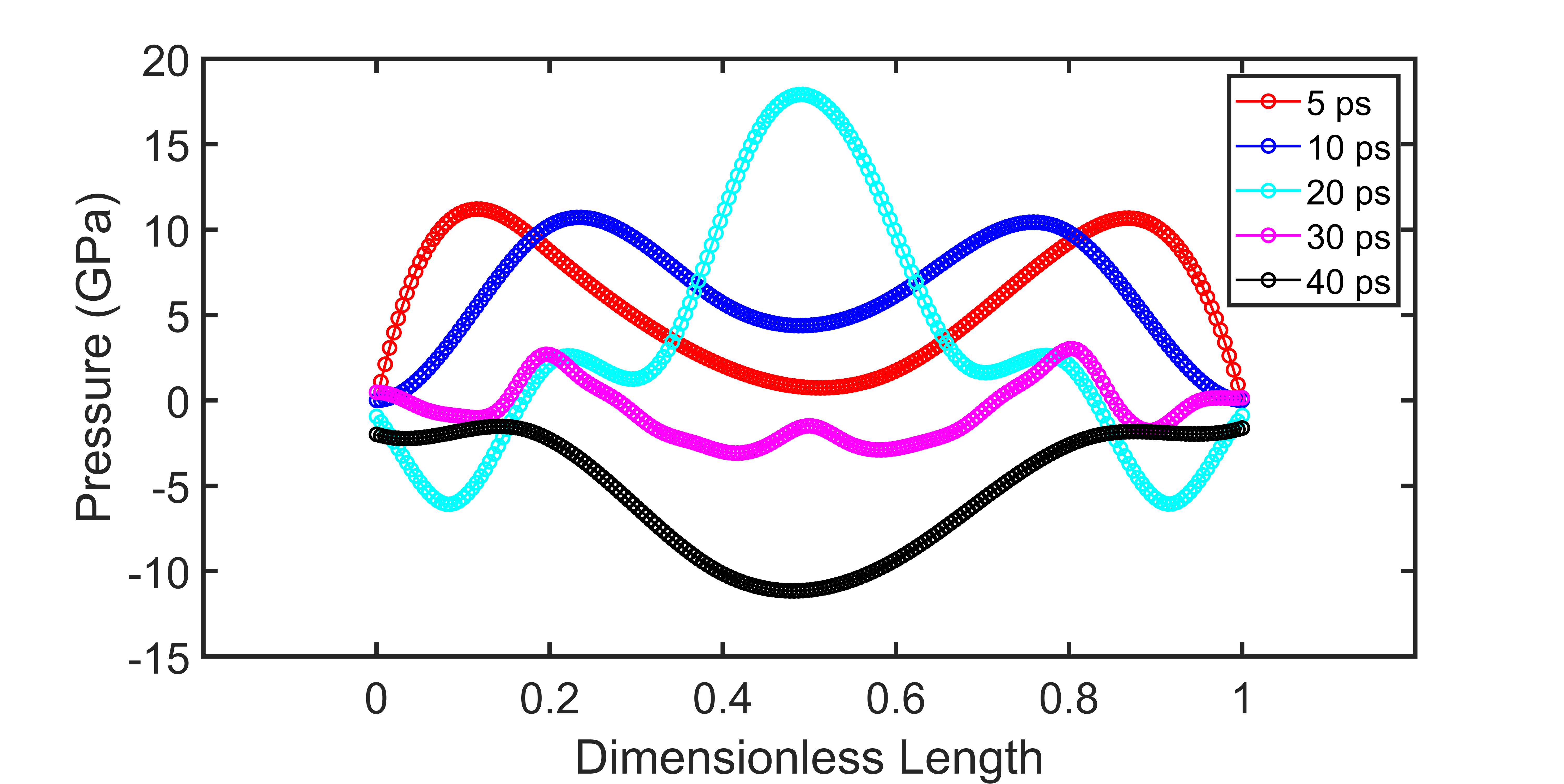}
	\caption{Shock profiles along the thickness with the sample during two-laser shock loading with a fluence of 1,250 $\mathrm{J\cdot m^{-2}}$ and a pulse duration of $t_{p}$ = 0.1 ps.}
	\label{fig:Pressure_profile_length}
\end{figure}

The shock wave profile, pressure along the direction of the laser, is displayed in Figure \ref{fig:Pressure_profile_length} for different times. 
As opposed to the shock response under plate impact simulations \cite{qiu2017non}, it can be seen that the profile is marked by a rapid attenuation, i.e., it lacks any constant zones behind the shock-front, and unlike in single-pulse laser spallation \cite{galitskiy2018dynamic}, two similar pulses propagate away from the front and back surfaces.
We observed that a large initial pressure develops early (i.e., 5 -10 ps). As the shock waves propagate, they eventually meet in the middle of the target, reaching the maximum pressure value ($t = 20$ ps). However, we also observed that a release wave is emitted behind the pressure front due to the free surface condition, as it is graphically evident in  Figure \ref{fig:Pressure_profile_length} for $t = 20$ ps. The interaction between the two waves becomes quite complex until the two release waves meet in the middle of the sample, where the stress state is characterized by a large hydrostatic stress value ($t = 40$ ps). After this point, the sample nucleated voids, which quickly grew and coalesced to generate a crack in the spall plane. 
Due to the reduced laser intensity requirement in the two-laser spall approach, the crystal structure does not undergo a phase transformation due to pressure, as observed in Figure \ref{fig:Pressure_profile_length} since the shock profile does not change in shape, thus indicating minimal FCC to BCC transformation and therefore reduced impact on the spall behavior \cite{galitskiy2018dynamic}.

\begin{figure}[h]
	\centering
	\includegraphics[width=1.0\textwidth]{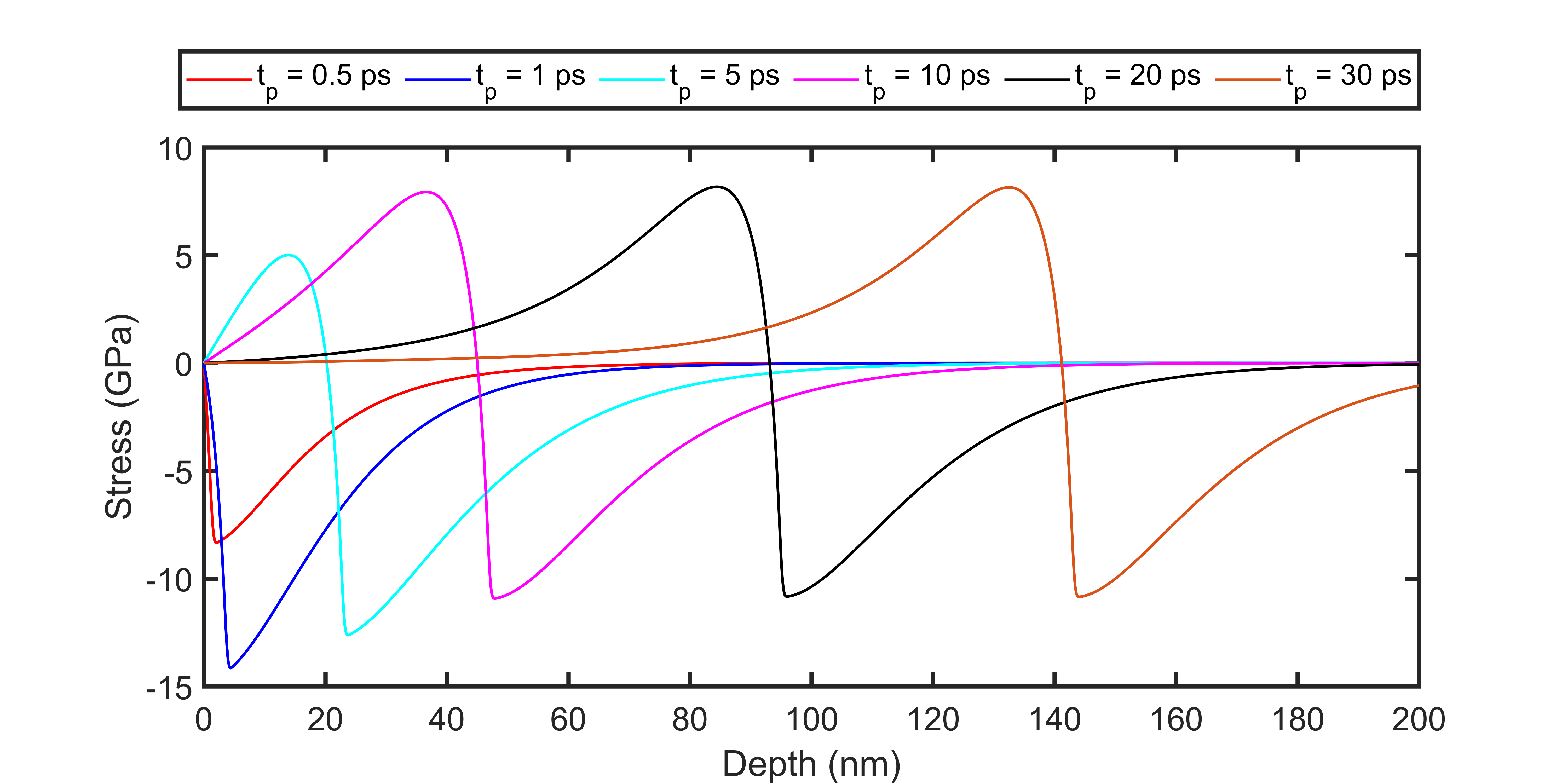}
	\caption{Analytically calculation of stress as a function of time as a result of thermoelastic effect with a fluence of 1,250 $\mathrm{J\cdot m^{-2}}$ and a pulse duration of $t_{p}$ = 0.1 ps.}
	\label{fig:Analytical_stress_profile_length}
\end{figure}

To interpret the shock wave profile and describe the compression-tension process, the propagation of the stress generated by a thermoelastic wave has been analytically calculated for nickel without considering any material melting/evaporation, phase transformation, and corresponding structural change \cite{wang2002thermoelastic}. 
According to the analytical solution in Figure \ref{fig:Analytical_stress_profile_length}, a single laser pulse with a duration of $t_p$ = 0.1 ps and a fluence of 1,250 $\mathrm{J \cdot m^{-2}}$ generates a tensile stress of the order of $\sim$ 8 GPa; this would result in a peak tensile stress of $\sim$ 16 GPa under the two-pulse laser configuration.
As the laser interacts with the surface of specimen, the heated region rapidly expands leading to the release of a compressive wave towards the interior of the target. 
However, as the atoms closest to the heated region cannot freely move, a complex interaction occurs, resulting in the release of unloading tensile waves as calculated in both MD (Figure \ref{fig:Pressure_profile_length}) and analytical solution (Figure \ref{fig:Analytical_stress_profile_length}).
This behavior was also observed in experiments \cite{romain2002pressure}, and agrees well with the idealized shock-wave dynamics in Figure \ref{fig:Idealized_shock_wave_behaviour_single_two_pulse_laser_approach}.

\begin{figure}[h!]
	\centering
	\begin{subfigure}{0.45\textwidth} \centering
		\includegraphics[width=0.99\textwidth]{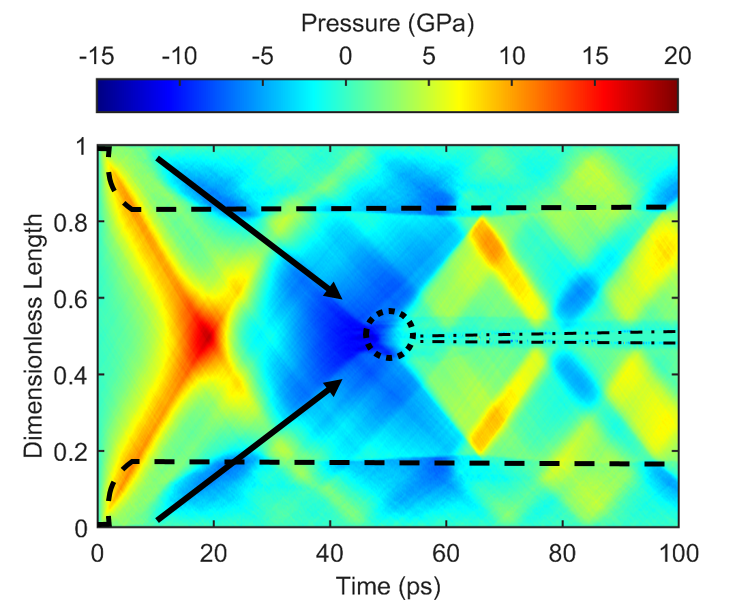}
		\caption{}
		\label{fig:Pressure_map_1250}
	\end{subfigure}
	\begin{subfigure}{0.45\textwidth} \centering
		\centering
		\includegraphics[width=0.99\textwidth]{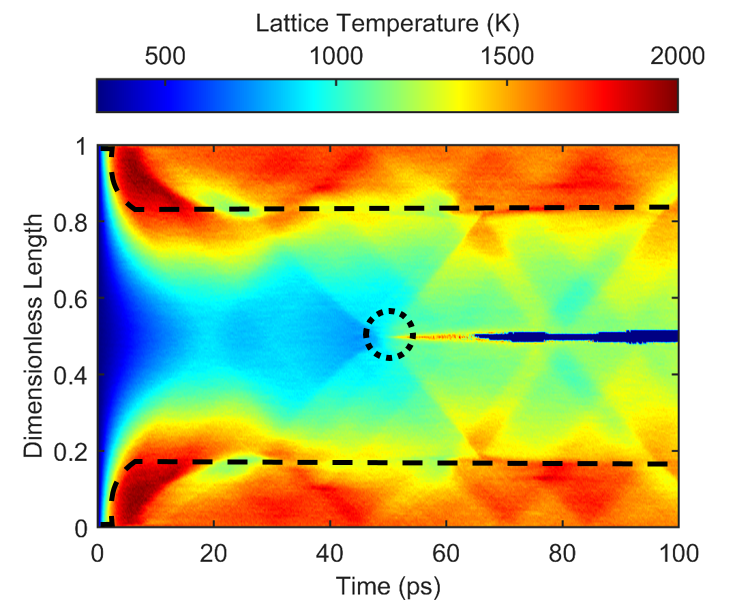}
		\caption{}
		\label{fig:Temperature_map_1250}
	\end{subfigure}
	\begin{subfigure}{0.8\textwidth} \centering
		\centering
		\includegraphics[width=0.99\textwidth]{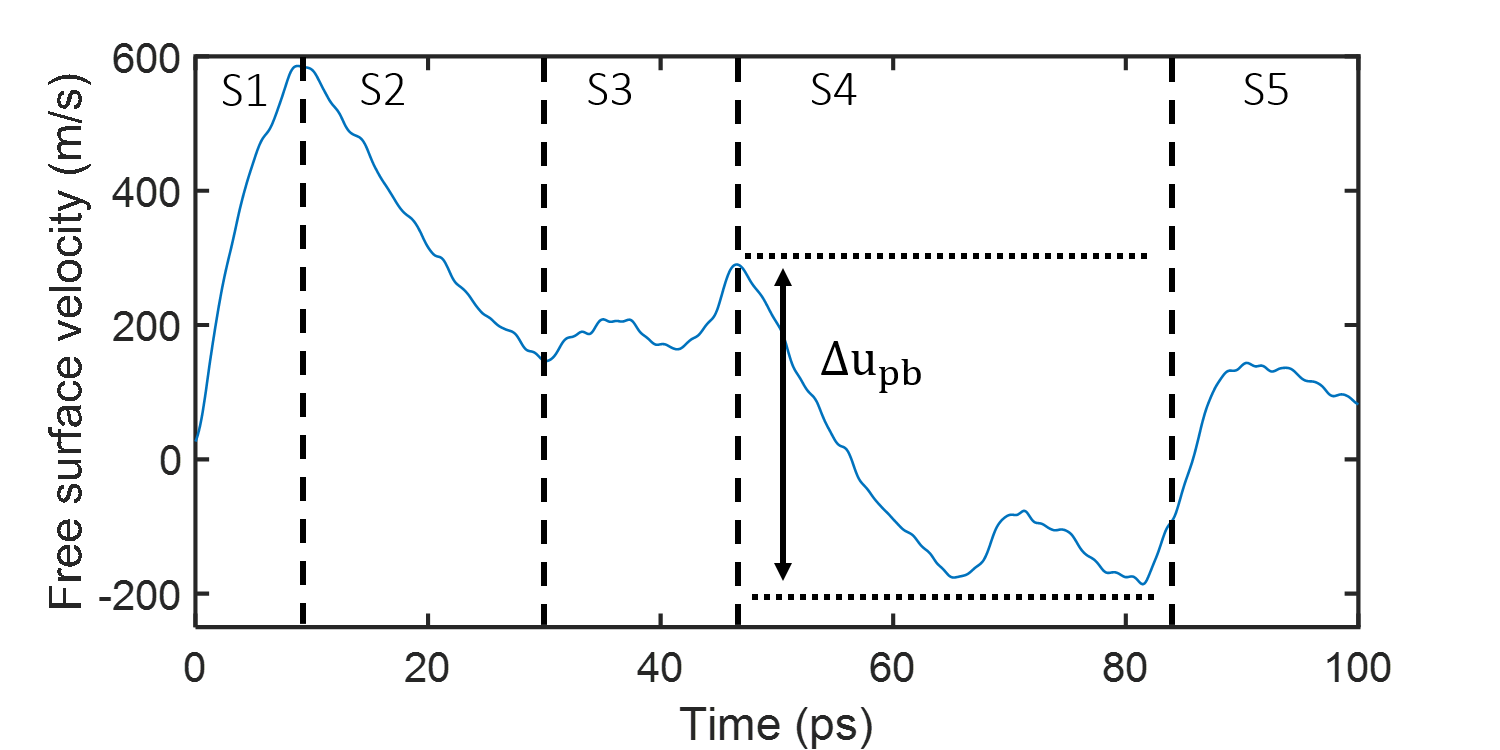}
		\caption{}
		\label{fig:free_surface_velocity_1250}
	\end{subfigure}
	\caption{Schematic of (a) the pressure and (b) temperature profile as a function of space and time for a laser fluence of 1,250 $\mathrm{J}\cdot \mathrm{m^{-2}}$ and a pulse duration of $t_{p}$ = 0.1 ps. The dashed lines represent the evolution of the solid-liquid interface. The two black arrows mark the interaction of the two unloading tensile waves. The dashed circle symbolizes the area of initial void nucleation. (c) displays the free surface velocity \emph{vs.} time when the spall signal can be identified.}
	\label{fig:pressure_temperature_free_surface_velocity_1250}
\end{figure}

The pressure evolution due to the interaction of the two-laser pulses with the sample as a function of space and time is displayed in Figure \ref{fig:Pressure_map_1250} using the 1D binning analysis. In the figure, the red regions represent significant compressive stresses, and the blue regions the tensile stresses.   
The rapid deposition of energy leads to the compressive pressure buildup as a result of the condition of stress confinement \cite{zhigilei2000microscopic}, which, in turn, relaxes by driving the compressive stress wave away towards the other end.
Due to the use of two laser pulses at the free ends of the specimen, the compressive shock load is replicated. 
Subsequently, the relaxation of the compressive pressure near the loaded surfaces leads to the release of unloading tensile waves of the pressure wave (solid arrows in Figure \ref{fig:Pressure_map_1250}).
The interaction of these two unloading tensile waves with each other generates a state of large hydrostatic tensile stress located at the center of the specimen.
Void nucleation is observed in the regions where the hydrostatic tensile stress has peaked, indicated with the dashed circle in Figure \ref{fig:Pressure_map_1250}. 
Subsequently, void growth and coalescence eventually occur to create new surfaces and initiate the spall failure of the material.
Void growth, as indicated by the dash-dot lines in Figure \ref{fig:Pressure_map_1250}, initiates at $\sim$50 ps, and the damaged area induced by the two laser pulses expands, leading to the creation of a new, traction-free surface.
The maximum hydrostatic tensile stress-recorded in the simulation via the binning analysis of the virial stress-arising from the interaction of these waves is used to determine the spall strength of a material.
The calculated maximum tensile stress is $\sim 12$ GPa. 

The lattice temperature evolution due to the laser energy deposition as a function of space and time is plotted in Figure \ref{fig:Temperature_map_1250}.
Laser-induced excitation of the conduction band electrons and rapid energy transference to the phonons leads to increased initial lattice temperature at the regions close to the irradiated surfaces.
The strong electron-phonon coupling in Ni results in a sharp temperature gradient, causing the temperature to rise to $\sim$1800 K at around 5 ps.
As shown in Figure \ref{fig:Temperature_map_1250}, the solid-liquid interface is restricted to the regions close to the loaded surfaces, and that void nucleation takes place in the solid region, well below the melting temperature $\sim$1730 K for Ni \cite{cezairliyan1984melting}.
Figure \ref{fig:free_surface_velocity_1250} shows the free surface velocity profile for one of the free surfaces of the specimen. 
The spall signal can be obtained allowing for the calculation of the spall strength ($\sigma_\mathrm{spall} = \frac{1}{2} \rho c_0 \Delta u_{pb}$) and strain rate ($\dot{e} = \frac{1}{2} \frac{\Delta u_{pb}}{c_0 \Delta t}$) using simple acoustic relationships \cite{grady1988spall}.
The spall strength calculated from Figure \ref{fig:free_surface_velocity_1250} is $\sim13.3$ GPa with a strain rate of $2.9\times 10^9 ~\mathrm{s}^{-1}$, in good agreement with the value obtained from the pressure profile obtained in the binning analysis, as depicted in Figure \ref{fig:Temperature_map_1250}. 
Similarly, the computed MD spall strength agrees very well with the value of $\sim13.2$ GPa extrapolated from the literature under similar strain rate levels \cite{srinivasan2007atomistic, golubev1983character, golubev1985effect, golubev1985nature}.
Moreover, the spall strength obtained from the MD simulations under the two-pulse approach is in good agreement with the experimentally calculated value for the microscale-thick Ni foil under the single-pulse approach ($\sim$ 12.9 GPa), despite the differences in thickness and strain rate.

The different stages of spallation under the two-laser configuration can be observed from the pressure, temperature, and free surface velocity profile.
The first stage is characterized by the melting of the region close to the irradiated surfaces resulting in the development of compressive shock fronts that propagate towards the center of the sample leading to a sharp increase in free surface velocity.
In the second stage, the unloading tensile waves are released, reducing the free surface velocity.
The third stage involves the interaction of tensile waves leading to the generation of voids at the spall plane.
Simultaneously, during this stage, the compressive waves reach the ends of the sample.
The spall signal can be traced in the fourth stage, marked by free surface deceleration, and corresponds to the classical pullback velocity signal.
Finally, stage five represents the reflection of waves away from the free surfaces.
\subsection{Molecular dynamics analysis of spall failure mechanism} 
Here, the dislocation behavior, void nucleation, growth, and coalescence are investigated, as shown in Figure \ref{fig:Schematic_MD_simulation}.
The results were obtained under a laser fluence of 1,250 $\mathrm{J}\cdot \mathrm{m^{-2}}$, with a pulse duration $t_{p}$ = 0.1 ps, and an initial temperature of 300 K.
To describe the failure analysis, we investigate two cases, perfect and imperfect crystals.
For the latter, a nanovoid of diameter 4 nm was generated at the center of the imperfect structure by removing atoms from the bulk crystal.
\subsubsection{Perfect Ni crystal}
\begin{figure}[h!]
	\centering
	\begin{subfigure}[b]{0.45\textwidth}
		\centering
		\includegraphics[width=\textwidth]{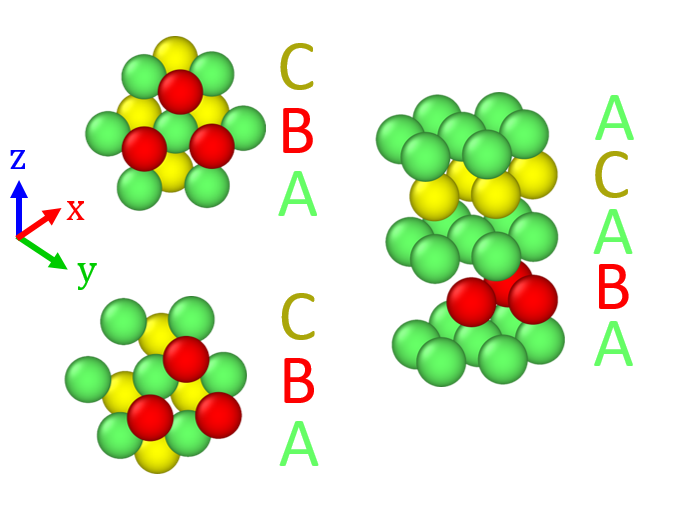}
		\caption{}
		\label{fig:Initial_no_void_stacking_layer_shift}
	\end{subfigure}
	\begin{subfigure}[b]{0.45\textwidth}
		\centering
		\includegraphics[width=\textwidth]{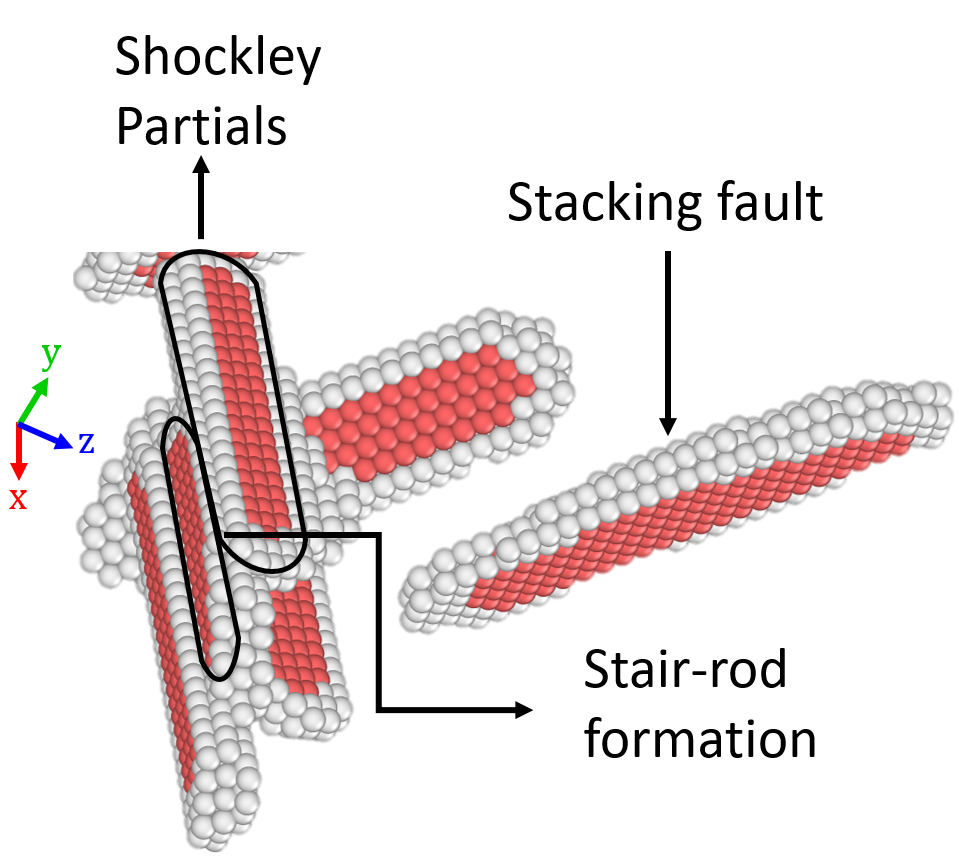}
		\caption{}
		\label{fig:stacking_fault_no_void}
	\end{subfigure}
	\caption{Atomistic simulation of the initial defect formation under the two-pulse laser spallation of a single perfect crystal Ni. Schematic (a) illustrates the FCC stacking sequence before (top) and after slipping (bottom), and the 4H-HCP structure (right) where the stacking layers A, B, and C are colored accordingly. The developed stacking faults as a result of atoms slipping at different closed packed planes, after removing the FCC atoms, can be seen in (b) where white, and red represent defect, and HCP atoms, respectively.}
	\label{fig:initial_perfect}
\end{figure}
During the initial stages of loading, the system behaves elastically, stretching the lattices without forming dislocations.
The coordination number of an undeformed FCC crystal is 12 and follows the stacking sequence of ...ABCABCABC... as shown in the first image in Figure \ref{fig:Initial_no_void_stacking_layer_shift}, however, as the tensile stress intensifies (when the two unloading tensile waves meet), certain atoms depart from their equilibrium lattice positions and tend to slip on the closed packed \hkl{111} planes along \hkl<110> directions.
Interestingly, it can be seen in Figure \ref{fig:Initial_no_void_stacking_layer_shift} that as a result of the shift in stacking layers, a 4H-HCP phase was generated with a stacking sequence of ...ABACA... \cite{orhan2021surface}.
As the atoms re-orient their positions on the closed packed planes, they interact with neighboring atoms leading to an increase in slipping between atoms and forming more stacking faults near the zone where the spall plane will later form \cite{pang2014dislocation}.
The developed stacking faults are oriented along the \hkl{111} planes and can be seen in Figure \ref{fig:stacking_fault_no_void}, where only faulted atoms are shown near the central zone of the simulation cell. 
Analyzing these defects reveals the presence of Shockley partials that develop along the slip directions and are parallel to the plane of fault.

\begin{figure}[h!]
	\centering
	\begin{subfigure}{0.4\textwidth} \centering
		\includegraphics[width=0.99\textwidth]{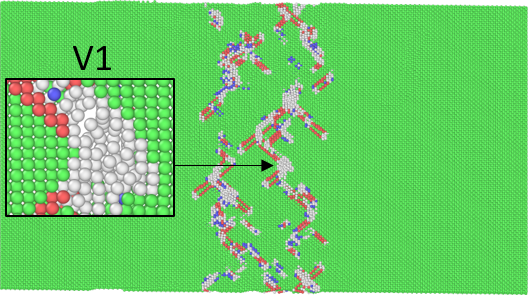}
		\caption{45 ps}
		\label{fig:microstructure_no_void_45_ps}
	\end{subfigure}
	\begin{subfigure}{0.4\textwidth} \centering
		\centering
		\includegraphics[width=0.99\textwidth]{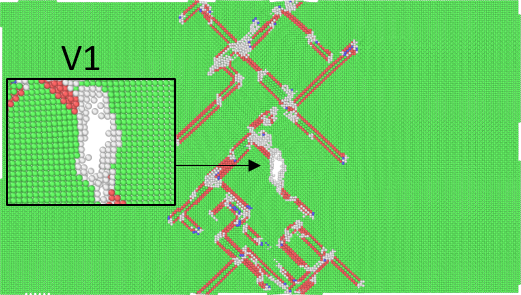}
		\caption{50 ps}
		\label{fig:microstructure_no_void_50_ps}
	\end{subfigure}
	\begin{subfigure}{0.4\textwidth} \centering
		\centering
		\includegraphics[width=0.99\textwidth]{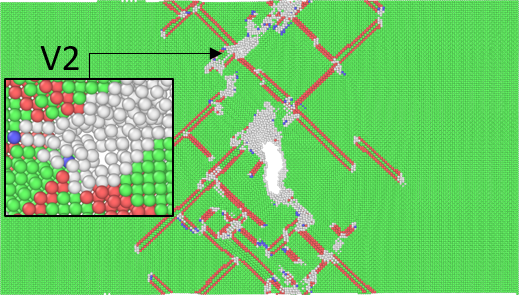}
		\caption{55 ps}
		\label{fig:microstructure_no_void_55_ps}
	\end{subfigure}
	\begin{subfigure}{0.4\textwidth} \centering
		\centering
		\includegraphics[width=0.99\textwidth]{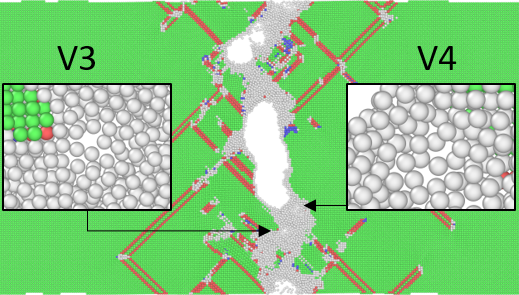}
		\caption{60 ps}
		\label{fig:microstructure_no_void_60_ps}
	\end{subfigure}
	\begin{subfigure}{0.49\textwidth} \centering
		\centering
		\includegraphics[width=0.99\textwidth]{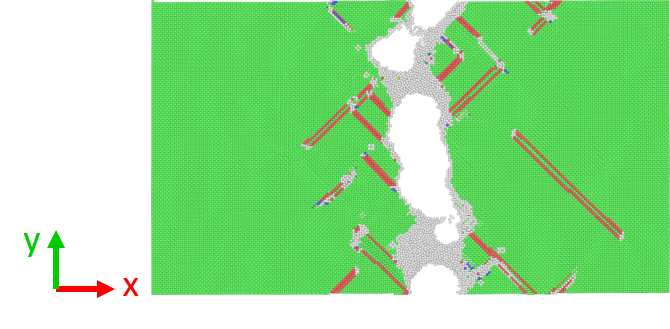}
		\caption{65 ps}
		\label{fig:microstructure_no_void_65_ps}
	\end{subfigure}
	\caption{Void nucleation, growth, and coalescence of a single perfect crystal Ni at the apex of interacting stacking faults under two-pulse laser spallation as a function of time between 45 ps to 65 ps, where white, red, blue, and green represents defect, HCP, BCC, and FCC atoms, respectively.}
	\label{fig:microstructure_perfect}
\end{figure}

More Shockley partials are emitted as the tensile state increases, leading to more stacking faults at the spall plane.
These Shockley partials trigger the onset of plasticity, and they glide and combine with dislocations moving on other intersecting \hkl{111} planes.
In this manner, stair-rod dislocations are formed along \hkl<110> directions, as shown in Figure \ref{fig:microstructure_no_void_45_ps}, where all atoms are now shown. Stair-rod dislocations are sessile and prevent the propagation of Shockley partials to the rest of the bulk \cite{he2019computational}.
Therefore, the interaction of these Shockley partials controls the spall plane location provided sufficient stress is generated to nucleate them \cite{gupta2016microstructure, dongare2011atomic}.
The emission of dislocations releases some deviatoric stress but not hydrostatic stress, accumulating a large amount of energy in the system as the tensile stress increases.
To release this energy, voids may nucleate at weak points in the material \cite{pang2014dislocation}.

\begin{figure}[h!]
	\centering
	\begin{subfigure}{0.32\textwidth} \centering
		\includegraphics[width=0.99\textwidth]{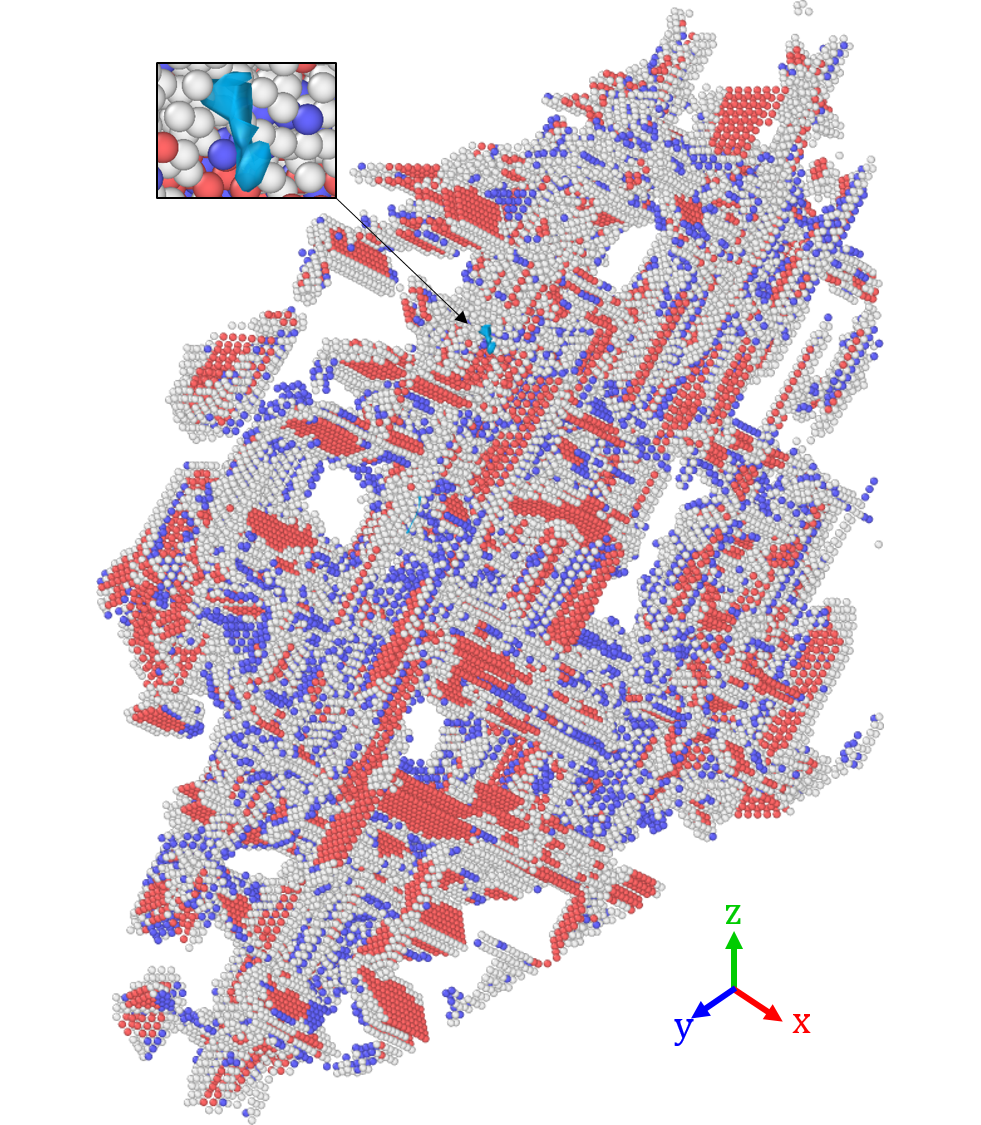}
		\caption{45 ps}
	\end{subfigure}
	\begin{subfigure}{0.32\textwidth} \centering
		\centering
		\includegraphics[width=0.99\textwidth]{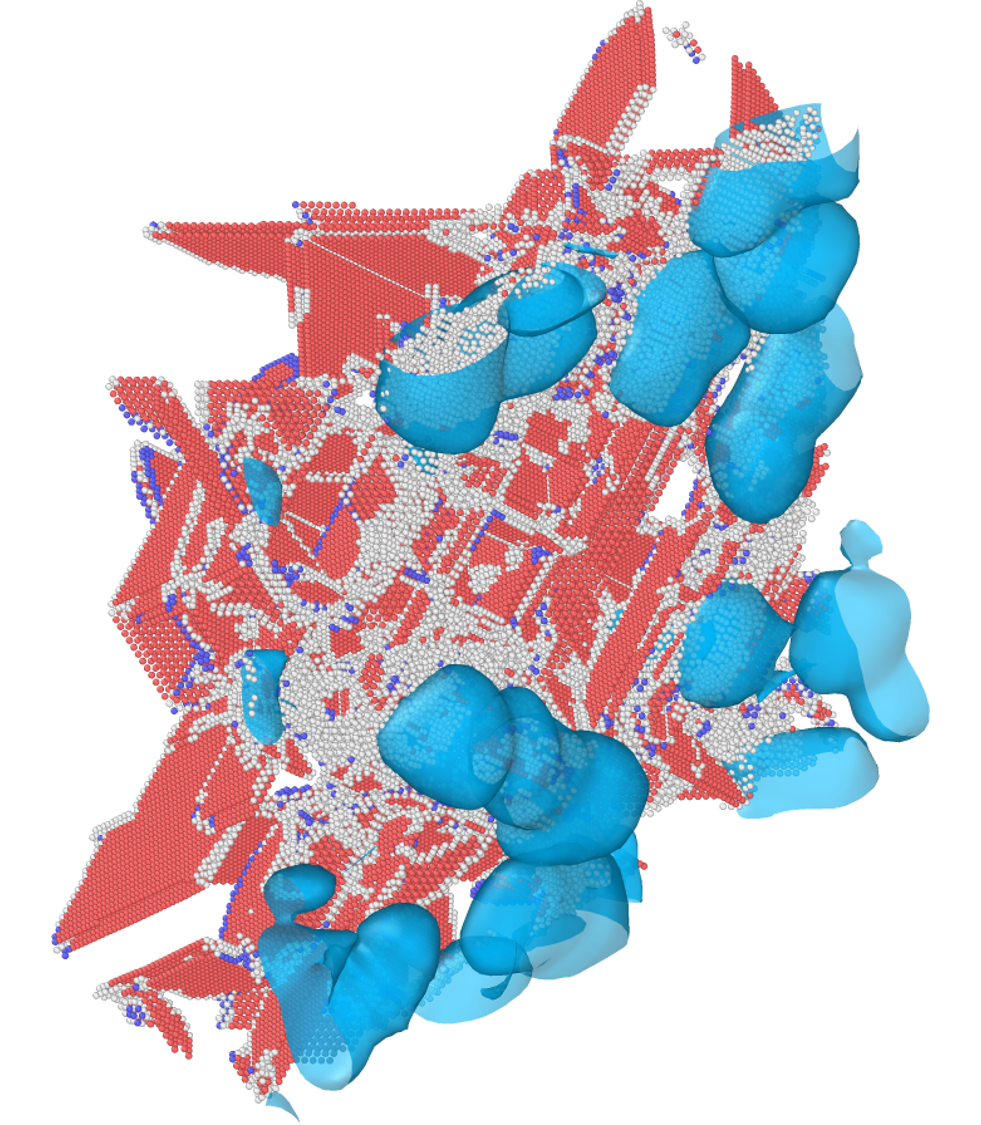}
		\caption{55 ps}
	\end{subfigure}
	\begin{subfigure}{0.32\textwidth} \centering
		\centering
		\includegraphics[width=0.99\textwidth]{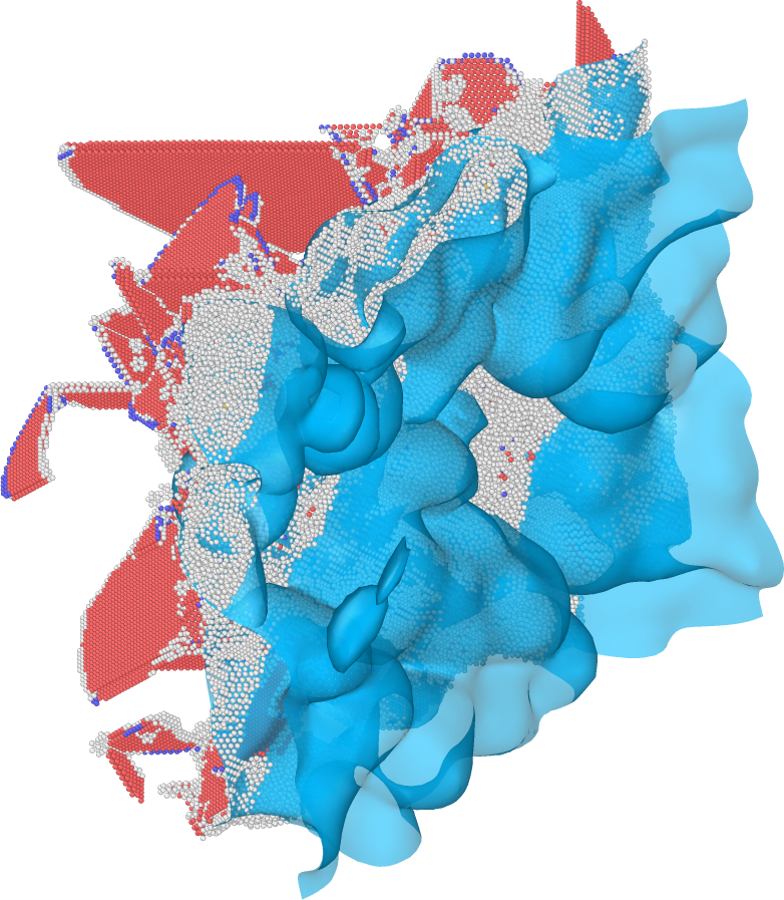}
		\caption{65 ps}
	\end{subfigure}
	\caption{Schematic of (a) void nucleation, (b) growth, and (c) coalescence of a single perfect crystal Ni under two-pulse laser spallation as a function of time from 45 ps to 65 ps where white, red, blue, and green represents defect, HCP, and BCC atoms, respectively.  One-half of the sample, along the direction of the laser, has been displayed here after removing all FCC atoms. Based on the alpha-shape method \cite{stukowski2012automated}, a surface mesh is superimposed on the emitted dislocation.}
	\label{fig:microstructure_full_perfect}
\end{figure}

Figure \ref{fig:microstructure_perfect} illustrates a series of 2D slices of the nucleation and growth of a void at an apex where stacking fault defects meet.
It can be seen here that the creation of stair-rod dislocations directly correlates with spall failure as they offer potential sites for the nucleation of voids \cite{galitskiy2018dynamic}.
The interaction of these stacking fault defects increases the number of sessile dislocations, leading to a coalescence of nanovoids.
It can be seen from Figure \ref{fig:microstructure_perfect}a-e that multiple voids (labeled V1, V2, V3, and V4) nucleate at the weak spots.
These voids are created by the shearing displacement of nearby atoms, leading to clusters of atoms in defective positions surrounding the voids (colored white according to the CNA analysis).
As deformation continues under the hydrostatic tensile state, the size of these voids tends to increase, leading to an increase in the number of defect atoms and a decrease in the FCC atoms around each void.
Finally, as the voids interact at different regions along the spall plane, they eventually coalesce, generating a free surface.
The creation of this new traction-free surface triggers the release of a wave back towards the free surface, resulting in a spall signal in the free surface velocity profile.
Figure \ref{fig:microstructure_full_perfect} displays the nucleation, growth, and coalescence of voids to visualize the failure better.
Interestingly, several nanovoids nucleated near the spall plane and expanded to generate a fully coalesced crack, leading to the spall plane. 

\begin{figure}[h!]
	\centering
	\includegraphics[width=1.0\textwidth]{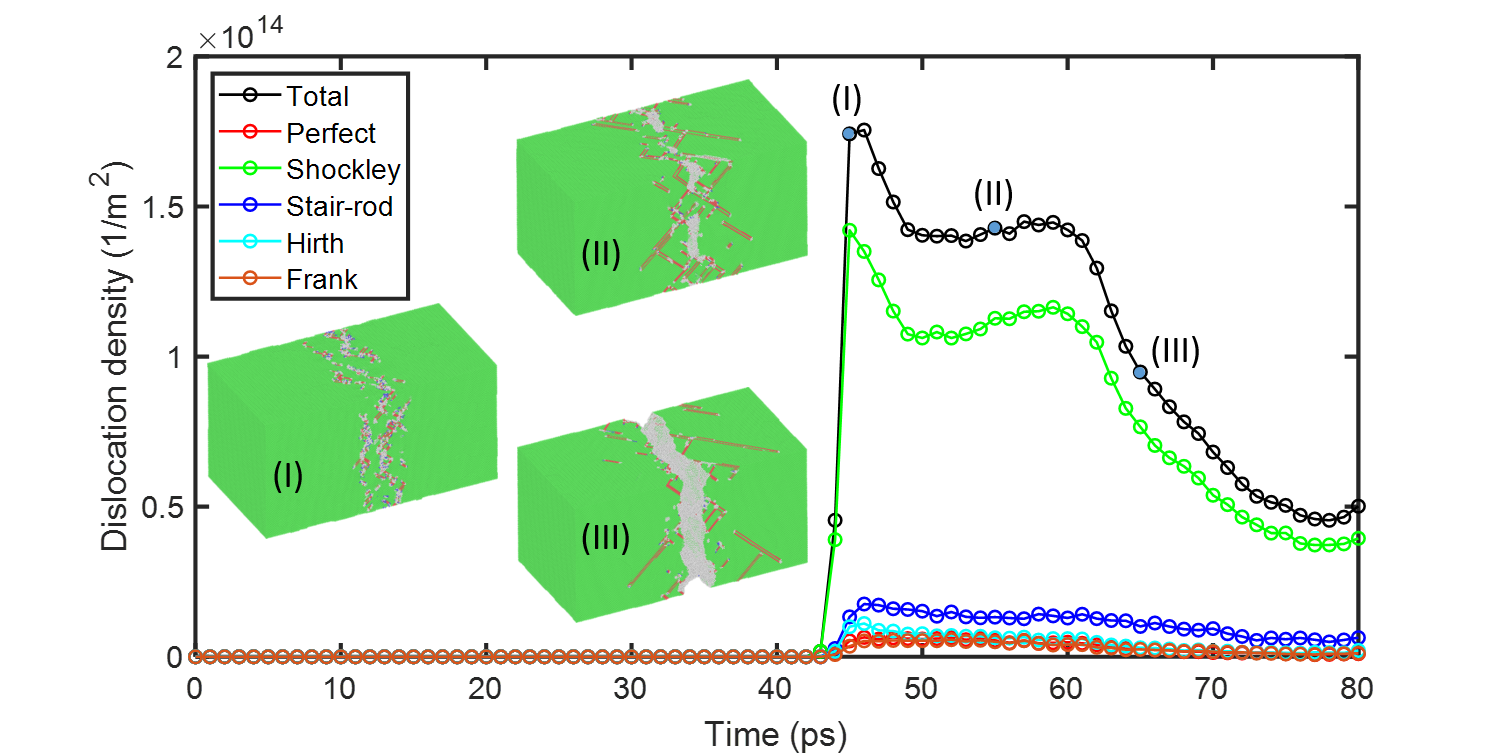}
	\caption{Temporal evolution of the dislocation density of a single perfect crystal Ni of 1,250 $\mathrm{J\cdot m^{-2}}$ and a pulse duration of $t_{p}$ = 0.1 ps. Snapshots of the deformed sample at 45, 55, and 65 ps correspond to the blue circles labelled with I, II, and III. }
	\label{fig:dislocation_pristine}
\end{figure}

Figure \ref{fig:dislocation_pristine} depicts the time evolution of the dislocation density in a perfect Ni crystal. 
The sample's deformation is observed to be dominated by Shockley partials.
A peak value of dislocation density is observed at about 45 ps, corresponding to the maximum tensile stress as shown in Figure \ref{fig:Pressure_map_1250}, resulting in void nucleation.
Following that, an increase in Shockley partial interactions leads to a decrease in the dislocation density (between 45 and 50 ps).
Subsequently, the growth of voids results in an increase in the density of Shockley partials (between 50 and 60 ps) which is then followed by the coalescence of voids (as shown in Figure \ref{fig:microstructure_perfect}).
A reduction of Shockley partials, at 60 ps, is observed as the tensile waves travel away from the spall region.
Furthermore, as compared to Shockley partials, the contributions of other dislocations to plasticity are found to be significantly lower; yet, as demonstrated, the presence of stair-rod enhances the likelihood for failure in perfect Ni crystal.
\subsubsection{Imperfect Ni crystal}
\begin{figure}[h!]
	\centering
	\begin{subfigure}[b]{0.32\textwidth}
		\centering
		\includegraphics[width=\textwidth]{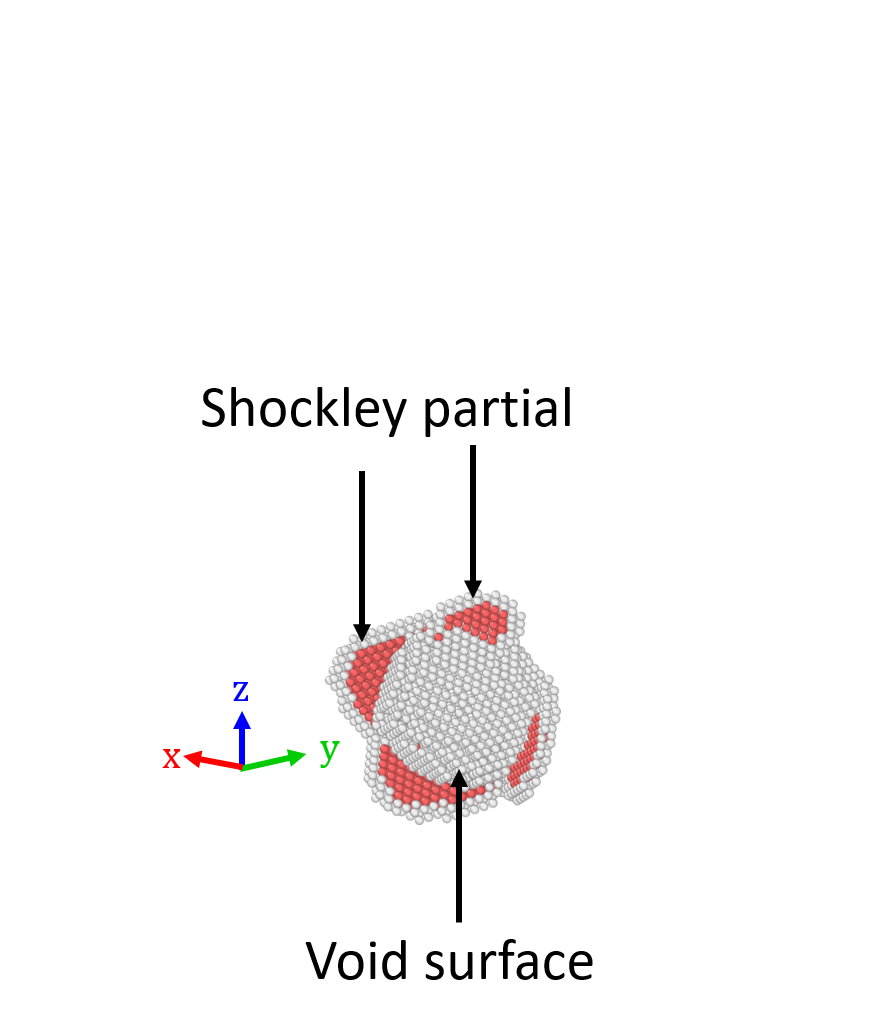}
		\caption{}
		\label{fig:Initial_void_dislocation_emission_1}
	\end{subfigure}
	\begin{subfigure}[b]{0.32\textwidth}
		\centering
		\includegraphics[width=\textwidth]{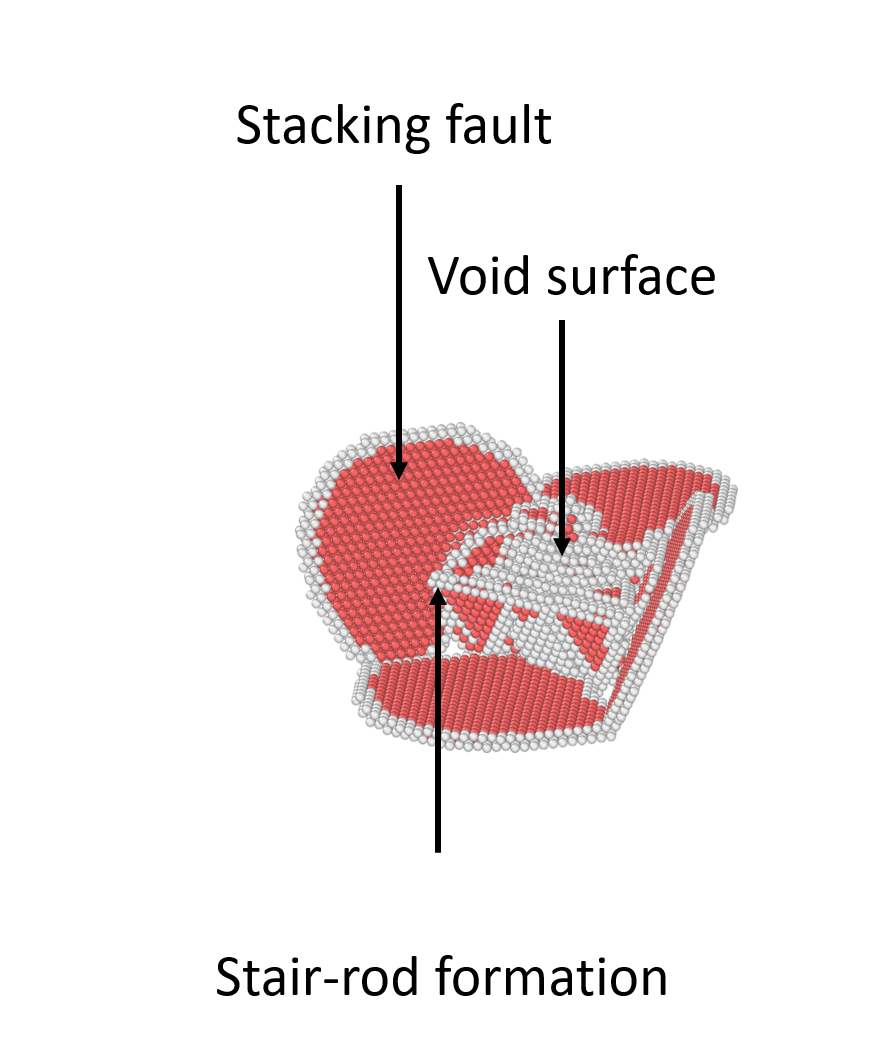}
		\caption{}
		\label{fig:Initial_void_dislocation_emission_2}
	\end{subfigure}
	\begin{subfigure}[b]{0.32\textwidth}
		\centering
		\includegraphics[width=\textwidth]{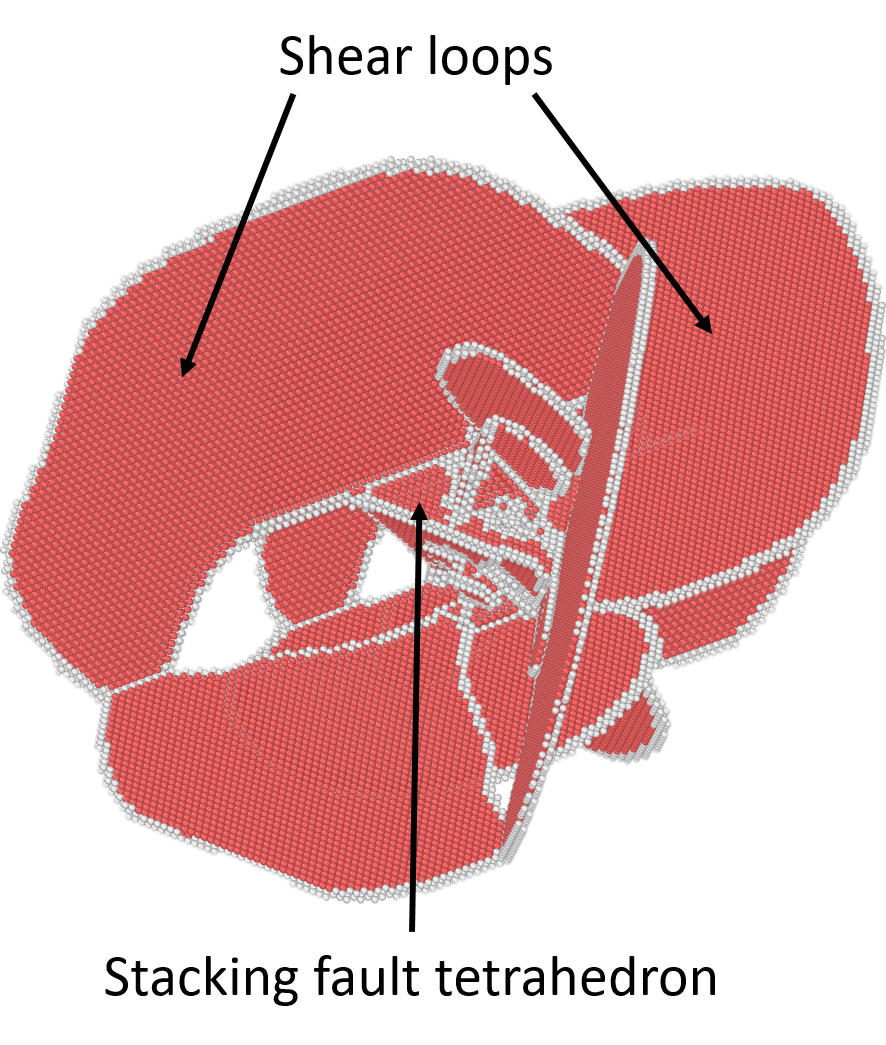}
		\caption{}
		\label{fig:Initial_void_dislocation_emission_3}
	\end{subfigure}
	\caption{Atomistic simulation of the initial defect formation under the two-pulse laser spallation of a single imperfect crystal Ni (with a 2 nm radius void). Snapshots (a) illustrate the nucleation of dislocation on the void surface, (b) the formation of stair-rod locks and (c) the development of stacking fault tetrahedron where white, and red represents defect, and HCP atoms, respectively.}
	\label{fig:initial_imperfect}
\end{figure}
Figure \ref{fig:initial_imperfect} depicts the nucleation and evolution of dislocations from the surface of the void under two-pulse laser loading.
In contrast to the perfect crystal, four Shockley partial dislocations are emitted during shock compression from the equator of the void along the closed-packed \hkl{111} planes.
At this stage in the process, sufficient shear stress is achieved between the closed-packed planes and the void surface, resulting in the emission of leading Shockley partials (as seen in Figure \ref{fig:Initial_void_dislocation_emission_1}).
The shear loop, in this case, comprises two layers of atoms parallel to the plane of the fault and separated by a stacking fault.
The leading and trailing Shockley partials in FCC metals are roughly structured as semi-circles, which are energetically attractive dislocation configurations due to the minimum stress needed \cite{chandra2018void,hirth_lothe_1982}.
As the load increases, additional shear loops are emitted from the void's surface on different \hkl{111} planes.
These secondary Shockley partials are highly mobile, gliding away from the void surface.
Eventually, these shear loops meet and intersect, forming stair-rod locks \cite{hirth_lothe_1982}, and an example reaction is given by $\frac{a}{6}[1 2 \overline{1}]+\frac{a}{6}[2 1 \overline{1}] \rightarrow \frac{a}{6}[\overline{1} 1 0]$.
The formation of these dislocations around voids has been observed in MD simulations \cite{ponga2015finite,prasad2021influence} and experiments \cite{epishin2016evolution,stevens1972spall}. 
These stair-rod locks form intersecting tetrahedrons around the void as depicted in Figure \ref{fig:Initial_void_dislocation_emission_2}.
Following the formation of stair-rod locks, more shear loops are nucleated from the unenclosed surface of the void as the presence of the tetrahedrons prevents the emission of any further dislocation.
Next, the stair-rod dislocation formed at the tetrahedron edge extends to accommodate the growth of shear loops leading to the formation of prismatic dislocation loops (as illustrated in Figure \ref{fig:Initial_void_dislocation_emission_3}) \cite{ponga2015finite}.
However, these shear loops annihilate with the passing of the compression waves.
\begin{figure}[h!]
	\centering
	\begin{subfigure}{0.19\textwidth} \centering
		\includegraphics[width=0.99\textwidth]{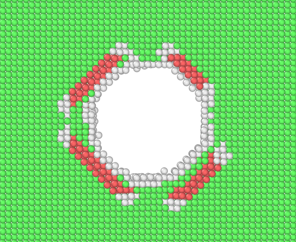}
		\caption{14 ps}
		\label{fig:void_collapse_1}
	\end{subfigure}
	\begin{subfigure}{0.19\textwidth} \centering
		\centering
		\includegraphics[width=0.99\textwidth]{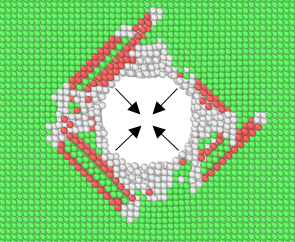}
		\caption{16 ps}
		\label{fig:void_collapse_2}
	\end{subfigure}
	\begin{subfigure}{0.19\textwidth} \centering
		\centering
		\includegraphics[width=0.99\textwidth]{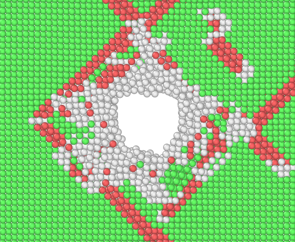}
		\caption{18 ps}
		\label{fig:void_collapse_3}
	\end{subfigure}
	\begin{subfigure}{0.19\textwidth} \centering
		\centering
		\includegraphics[width=0.99\textwidth]{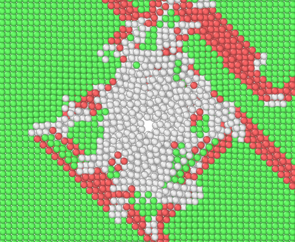}
		\caption{20 ps}
		\label{fig:void_collapse_4}
	\end{subfigure}
	\begin{subfigure}{0.19\textwidth} \centering
		\centering
		\includegraphics[width=0.99\textwidth]{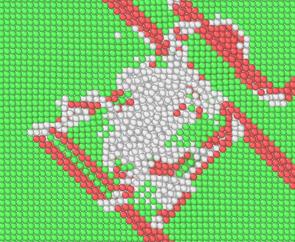}
		\caption{22 ps}
		\label{fig:void_collapse_5}
	\end{subfigure}
	\caption{Void collapse of a single imperfect crystal Ni (with a 2 nm radius void) from 14 ps to 22 ps due to shock compression where white, red, and green represent defect, HCP, and FCC atoms, respectively.}
	\label{fig:void_collapse}
\end{figure}

Figure \ref{fig:void_collapse} depicts the evolution of the initial void over time.
The initial void collapse is primarily governed by plasticity and internal jetting depending on the shock intensity \cite{xiang2017shock}.
It is well understood that plasticity is related to shear stress, whereas internal jetting is governed by hydrostatic pressure \cite{liao2018molecular}.
As was shown in Figure \ref{fig:initial_imperfect}, dislocation emission occurs on the void surface leading to the formation of shear loops along the \hkl{111} planes.
These shear loops shift the atoms on the void's surface and induce a transversal deformation.
Simultaneously, the void is in a state of hydrostatic (compression) pressure, producing a longitudinal deformation on the void.
Analyzing the void profile, it can be seen that the flow direction can be decomposed into a horizontal and vertical component, and this is due to the combined effect of plasticity and internal jetting.
Since the two compressive waves meet at the center of the sample, i.e., the location of the void, a symmetric-like deformation takes place, leading to the collapse of the void.
The collapsed void leaves behind vacancies and defects, which become prime sites for spall failure due to the tensile state.
\begin{figure}[h!]
	\centering
	\begin{subfigure}[b]{0.45\textwidth}
		\centering
		\includegraphics[width=\textwidth]{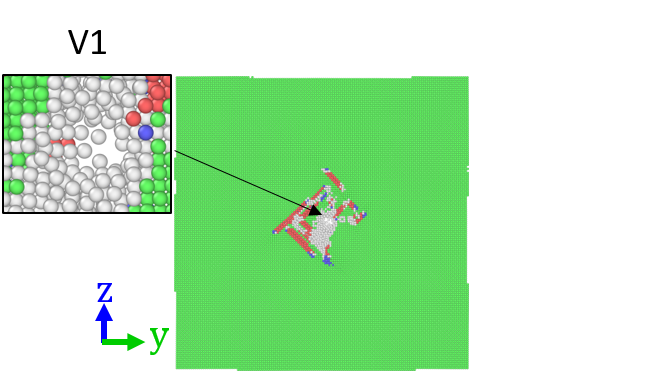}
		\caption{30 ps.}
		\label{fig:microstructure_void_30_ps}
	\end{subfigure}
	\begin{subfigure}[b]{0.45\textwidth}
		\centering
		\includegraphics[width=\textwidth]{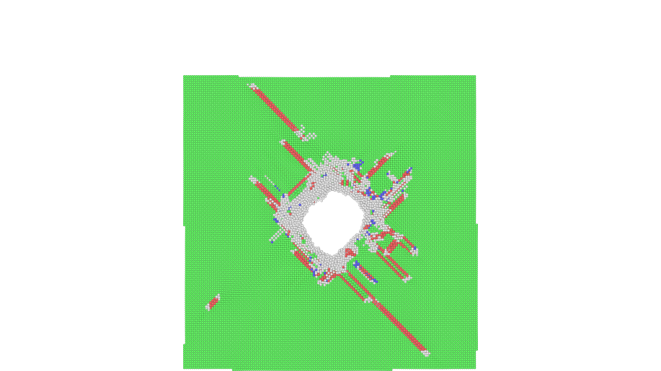}
		\caption{40 ps.}
		\label{fig:microstructure_void_40_ps}
	\end{subfigure} \\
	\begin{subfigure}[b]{0.45\textwidth}
		\centering
		\includegraphics[width=\textwidth]{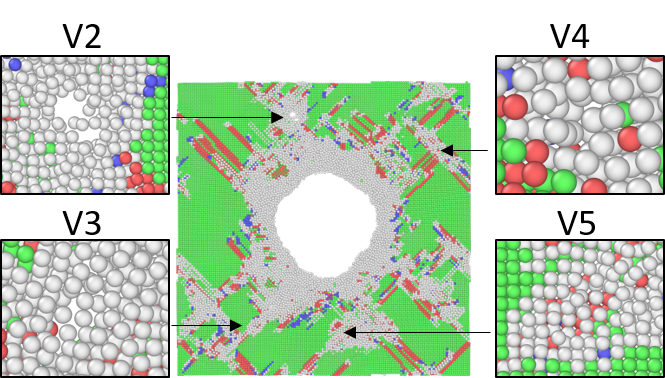}
		\caption{45 ps.}
		\label{fig:microstructure_void_45_ps}
	\end{subfigure}
	\begin{subfigure}[b]{0.45\textwidth}
		\centering
		\includegraphics[width=\textwidth]{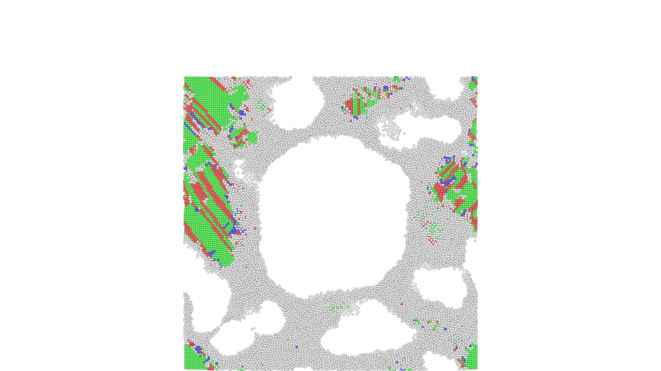}
		\caption{50 ps.}
		\label{fig:microstructure_void_50_ps}
	\end{subfigure}
	\caption{Void nucleation, growth, and coalescence of a single imperfect crystal Ni (with a 2 nm radius void) at the centre of the sample under two-pulse laser spallation as a function of time from 30 ps to 50 ps where white, red, blue, and green represent defect, HCP, BCC, and FCC atoms, respectively.}
	\label{fig:microstructure_imperfect}
\end{figure}

\begin{figure}[h!]
	\centering
	\begin{subfigure}{0.32\textwidth} \centering
		\includegraphics[width=0.99\textwidth]{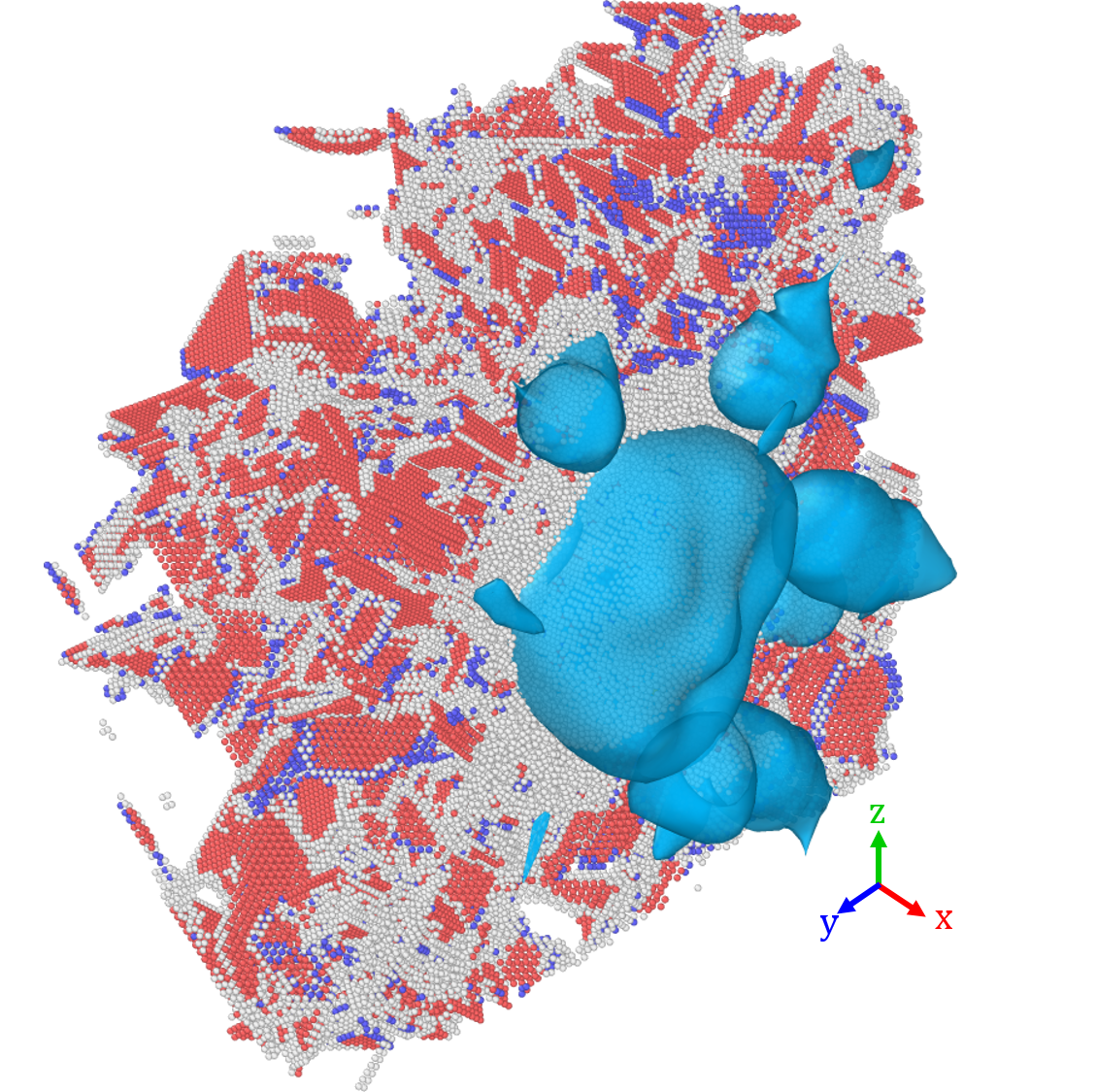}
		\caption{45 ps}
	\end{subfigure}
	\begin{subfigure}{0.32\textwidth} \centering
		\centering
		\includegraphics[width=0.99\textwidth]{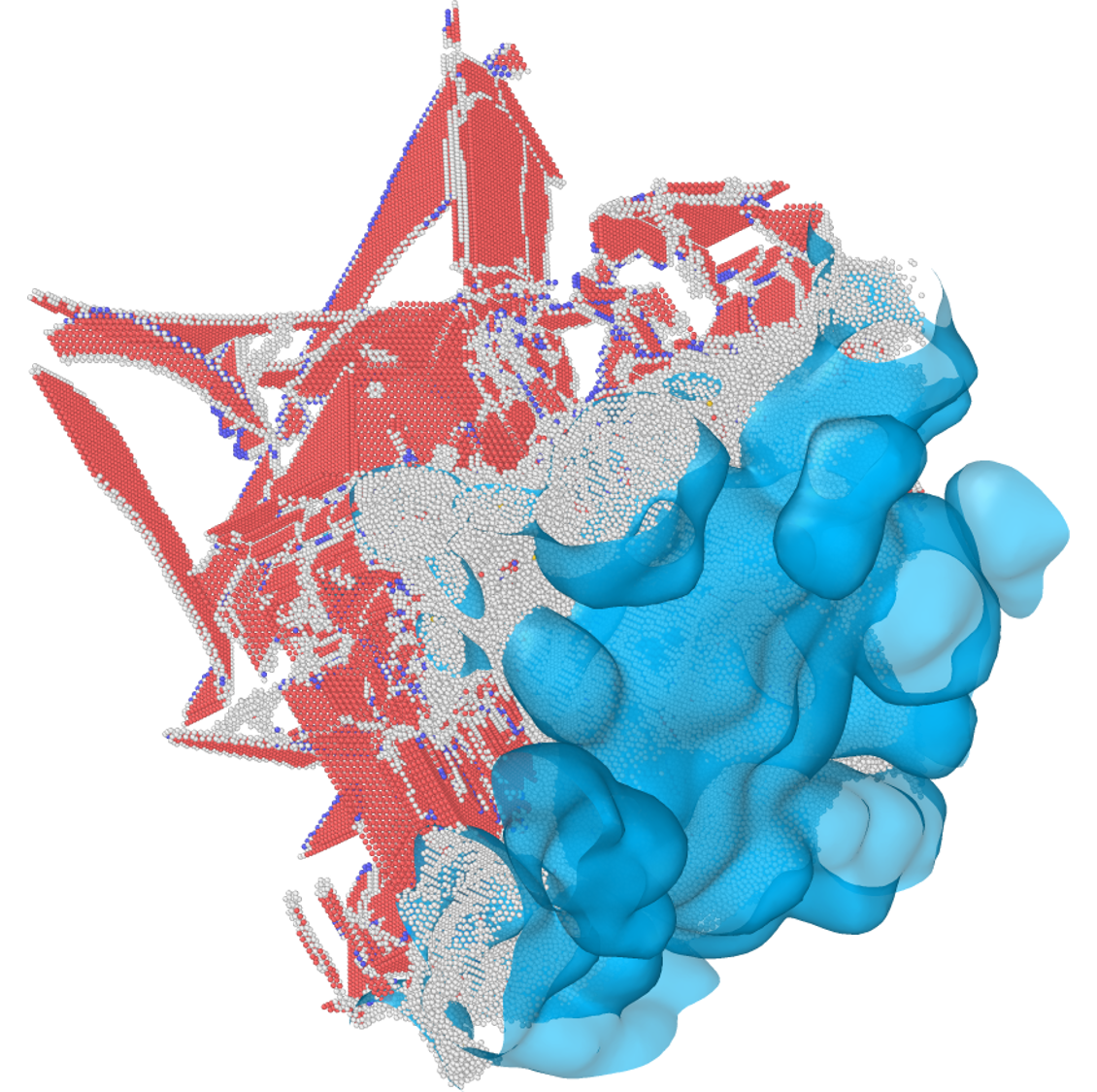}
		\caption{55 ps}
	\end{subfigure}
	\begin{subfigure}{0.32\textwidth} \centering
		\centering
		\includegraphics[width=0.99\textwidth]{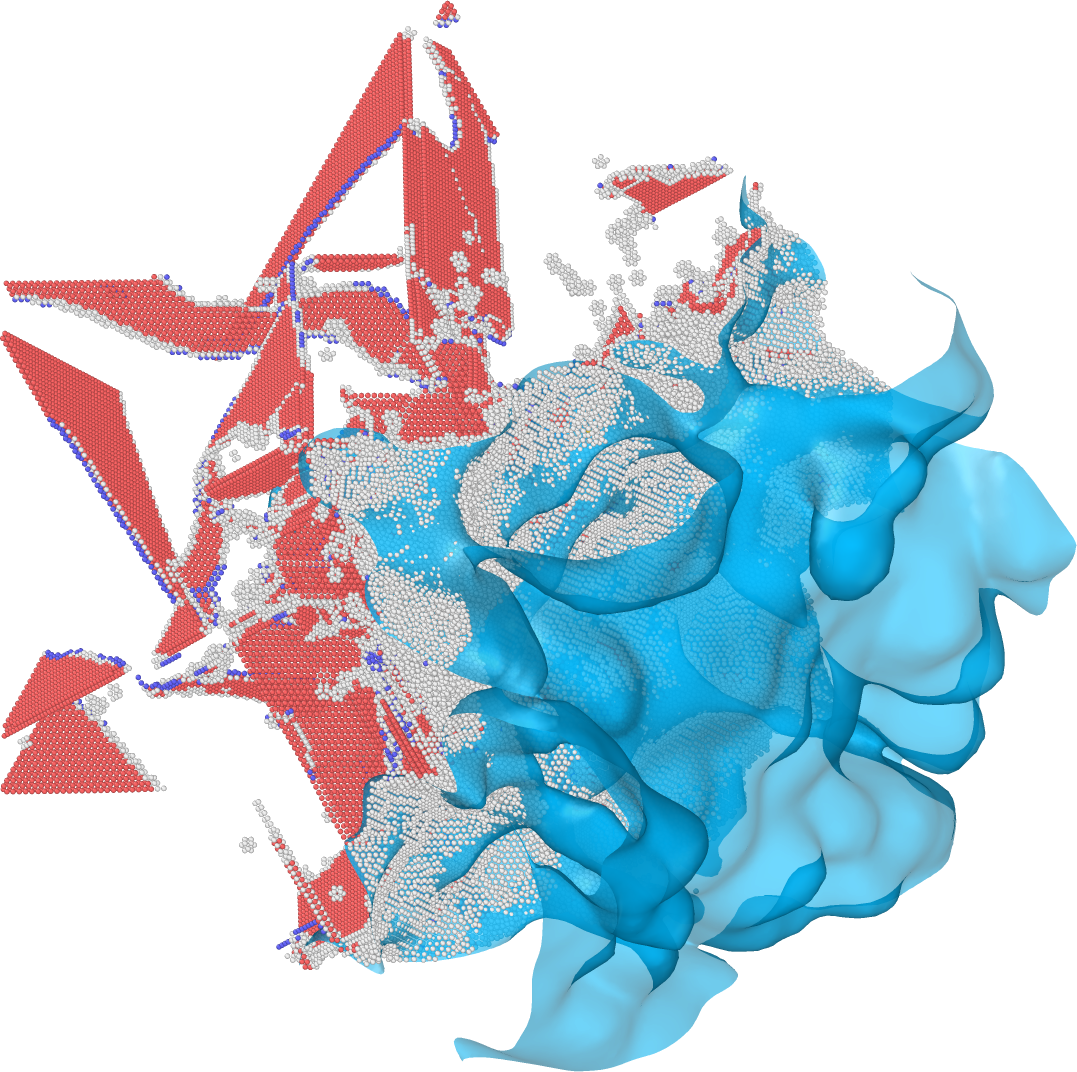}
		\caption{65 ps}
	\end{subfigure}
	\caption{Schematic of (a) void nucleation, (b) growth, and (c) coalescence of a single imperfect crystal Ni (with a 2 nm radius void) under two-pulse laser spallation as a function of time between 45 ps to 65 ps where white, red, blue, and green represents defect, HCP, and BCC atoms, respectively.  One-half of the sample, along the direction of the laser, has been displayed here after removing all FCC atoms. Based on the alpha-shape method \cite{stukowski2012automated}, a surface mesh is superimposed on the emitted dislocation.}
	\label{fig:microstructure_full_imperfect}
\end{figure}

Following shock compression, the two unloading tensile waves meet at the centre of the sample.
Figure \ref{fig:microstructure_imperfect} displays a series of 2D slices of the the nucleation and growth of a void at the collapsed initial void.
It can be seen here that as a result of shock compression on the initial void, vacancies have been left at the centre of the sample, and upon experiencing a hydrostatic tensile state, void nucleation takes place.
This is the driving mechanism behind spallation here, however, a similar mechanism to Figure \ref{fig:microstructure_perfect} takes place away from the collapsed void, where voids nucleates at intersections between shear loops \cite{ponga2015finite}.
The growth of the initial void (labelled as V1) takes the shape of an octahedra (as seen in Figure \ref{fig:microstructure_void_40_ps}), and this is consistent with experimental findings for single FCC crystals \cite{stevens1972spall,brown2015correlations}.
After which, multiple voids nucleate at different weak spots along the spall plane (labelled as V2, V3, V4, and V5).
As the atoms shear against each other, they deform leading to the increase in number of defect atoms surrounding the voids. 
As a result, as the load continues, the voids tend to grow, and eventually coalescence to generate a free surface as shown in Figure \ref{fig:microstructure_full_imperfect}.
It is evident that the void in the centre of the sample, originating from the collapsed void, expanded and interacts with voids nucleated elsewhere along the spall plane to eventually coalescence.

\begin{figure}[h!]
	\centering
	\includegraphics[width=1.0\textwidth]{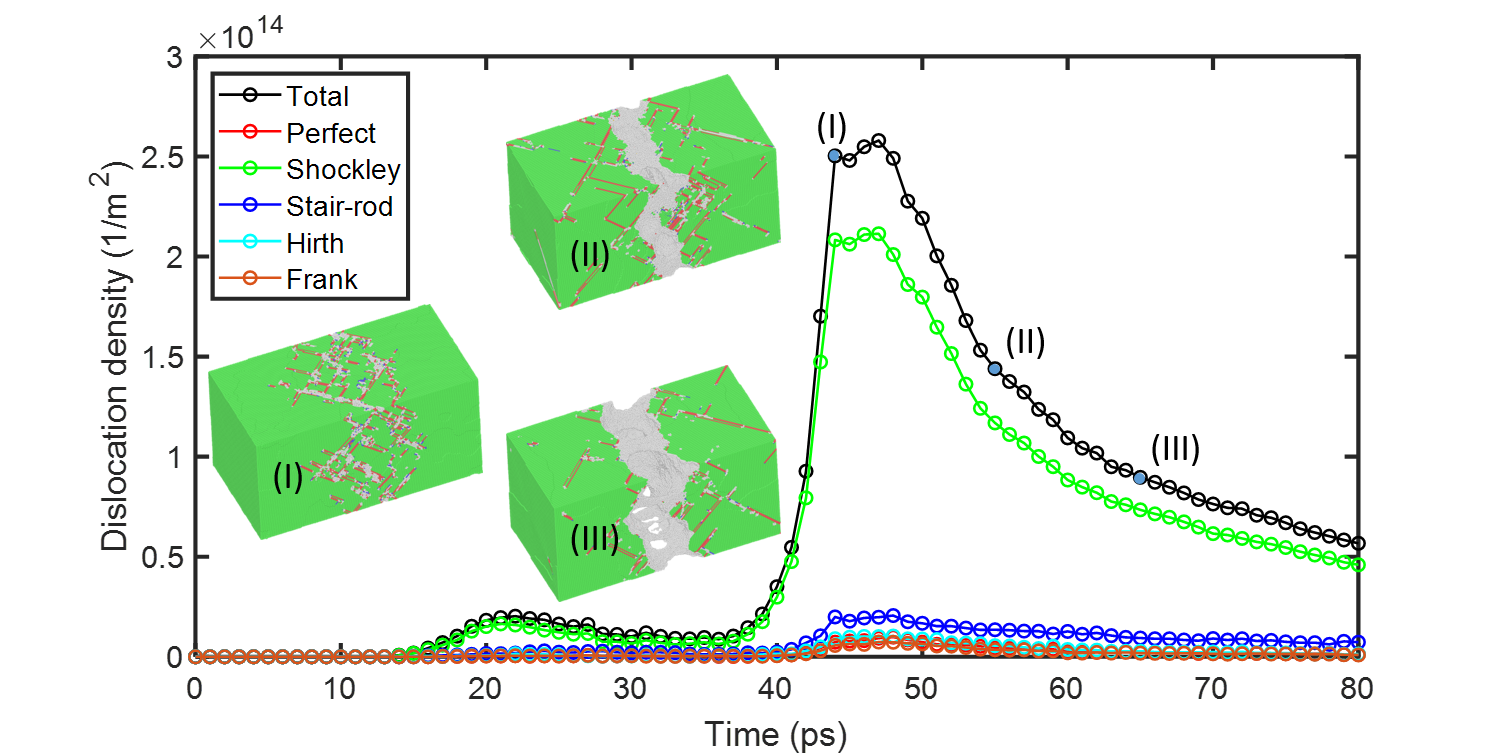}
	\caption{Temporal evolution of the dislocation density of a single imperfect crystal Ni (with a 2 nm radius void) of 1,250 $\mathrm{J\cdot m^{-2}}$ and a pulse duration of $t_{p}$ = 0.1 ps. Snapshots of the deformed sample at 45, 55, and 65 ps correspond to the blue circles labeled with I, II, and III. }
	\label{fig:dislocation_imperfect}
\end{figure}

Figure \ref{fig:dislocation_imperfect} depicts the evolution of dislocation density over time in an imperfect Ni crystal.
An initial peak is reached at around 20 ps due to the emission of dislocations during shock compression, where a state of maximum compression was experienced (see Figure \ref{fig:Pressure_profile_length}).
Following this, dislocations start to annihilate as the initial void collapses, leaving behind vacancies and defects.
Like Figure \ref{fig:dislocation_pristine}, a peak value corresponding to the maximum tensile stress is achieved at $\sim$45 ps. 
However, in this case, the peak value is nearly 40$\%$ higher.
In addition, when comparing the two cases (cf., Figures \ref{fig:microstructure_perfect} and \ref{fig:microstructure_imperfect}), it can be seen that upon reaching the peak dislocation density, void nucleation occurs due to the interaction of Shockley partials.
The critical difference is that the collapsed void opens before the nucleation of further voids elsewhere (see Figure \ref{fig:microstructure_void_40_ps}a and c), and hence, void coalescence occurs quicker in the imperfect case.
Therefore, a smooth decrease in the density of Shockley partials is seen after reaching the peak, as opposed to the plateau seen in Figure \ref{fig:dislocation_imperfect}.
The contributions from the other dislocations are minimal, where stair-rod dislocations indicate the likelihood of spall failure.
\section{Conclusion}
The effect of a novel femtosecond two-pulse laser spall approach was investigated experimentally using microscale- and nanoscale-thick nickel foils as a target material at strain rates on the order of $10^8$ to $10^9 ~\mathrm{s}^{-1}$, with a maximum laser pulse energy of 60 $\mu$J.
In this method, the front and back surfaces are illuminated by a single laser pulse at the same time.
When a secondary laser pulse is introduced in the two-pulse approach, an additional set of compressive and unloading tensile waves is generated as highlighted in Figure \ref{fig:Idealized_shock_wave_behaviour_single_two_pulse_laser_approach}. 
This study explored the potential impact introduced by the secondary laser pulse on spall failure.
The main findings of these systematic experiments, supported by atomistic simulations, are as follows:
\begin{itemize}
	\item Surface observations of the spalled microscale-thick Ni foil (see Figure \ref{fig:Surface_observation_single_two_pulse_laser_approach}) under the single- and two-pulse laser approach show similar characteristics. The key difference, however, is that the two-pulse laser approach produces spall features on both front and back surfaces such as an ablation crater, flaking, and an exposed spall plane. This difference can be understood by analyzing Figure \ref{fig:Idealized_shock_wave_behaviour_single_two_pulse_laser_approach}. In the single-pulse approach, spall occurs at location (I), whereas in the two-pulse approach, spall can occur at location (I) as well as (III) giving rise to spall features at both the front and back surfaces.
	\item Spall behavior observed in the microscale- and nanoscale-thick Ni foil (cf., Figures \ref{fig:Microscale_thick_Ni_spall_failure_observation_single_two_pulse_laser_approach} and \ref{fig:Nanoscale_thick_Ni_spall_failure_observation_single_two_pulse_laser_approach}) indicates that simultaneous illumination of the front and back surfaces can effectively induce failure at the center of the sample, away from the free surfaces. Achieving this outcome requires careful tuning of the laser fluence, laser alignment, and  considering wave attenuation to promote hydrostatic stress development at location (II) rather than locations (I) and (III) in Figure \ref{fig:Idealized_shock_wave_behaviour_two_pulse_laser_approach}. 
	\item MD analysis (see Figure \ref{fig:pressure_temperature_free_surface_velocity_1250}) revealed that, due to the low laser fluence, melting is limited to the regions near the loaded surfaces, and that spall failure does indeed occur in the solid state away from the free surfaces (as observed experimentally in Figure \ref{fig:Microscale_thick_Ni_spall_failure_observation_single_two_pulse_laser_approach}), due to the interaction of the unloading tensile waves. The calculated spall strength of 13.3 GPa from the MD produced free surface velocity agrees very well with the existing findings in the literature under similar levels of strain rate \cite{srinivasan2007atomistic, golubev1983character, golubev1985effect, golubev1985nature}. Moreover, the spall strength obtained from the MD simulations under the two-pulse approach is in good agreement with the experimentally calculated value for the microscale-thick Ni foil under the single-pulse approach ($\sim$ 12.9 GPa), despite the differences in thickness and strain rate.
	\item Spall failure analysis through MD highlighted that the interaction of stacking faults leads to the formation of sessile dislocations, at the spall plane, that serves as a prime location for the nucleation of voids with increasing tensile stress. Furthermore, when compared to a perfect crystal, the presence of a pre-existing void increases dislocation emission, resulting in the formation of more voids. The rapid expansion of these voids is followed by void coalescence, which results in the formation of a free surface.
\end{itemize}
Thus, our findings demonstrate that by illuminating both surfaces of a specimen, failure can be shifted from the free surfaces toward the center, replicating behavior typically observed in plate impact methods. 
This two-pulse approach requires a lower laser fluence than the single-pulse method, thereby reducing potential thermal effects and making it an attractive technique for high-throughput photo-mechanical experiments on the spall behavior of metals.
The authors are currently extending this approach to explore its application in studying the spall behavior of ceramics.
\section*{Code availability}
The codes developed to analytically calculate the thermoelastic stress and the $\ell2$T-MD model (extended to simulate the two-pulse laser approach) can be found at:
\begin{itemize}
	\item \url{https://github.com/mewael-isiet/analytical-thermoelastic-wave}
	\item \url{https://github.com/mewael-isiet/l2T-MD-two_pulse_laser}
\end{itemize}
\section*{CRediT author statement}
\textbf{Mewael Isiet}: Methodology, Software, Validation, Formal analysis, Investigation, Data Curation, Visualization, Writing - Original Draft, Writing - Review \& Editing. \textbf{Yunhuan Xiao}: Methodology, Writing - Review \& Editing. \textbf{Jerry I. Dadap}: Methodology, Supervision, Writing - Review \& Editing. \textbf{Ziliang Ye}: Resources, Writing - Review \& Editing. \textbf{Mauricio Ponga}: Conceptualization, Methodology, Software, Resources, Project administration, Funding acquisition, Supervision, Writing - Review \& Editing.
\section*{Data availability}
Data will be made available on request.
\section*{Acknowledgements}
We acknowledge the support of the New Frontiers in Research Fund (NFRFE-2019-01095) and the Natural Sciences and Engineering Research Council of Canada (NSERC) through the Discovery Grant and ALLRP 560447-2020 grants. This research was supported through high-performance computational resources and services provided by Advanced Research Computing at the University of British Columbia and the Digital Research Alliance of Canada. We acknowledge the support of Canada Foundation for Innovation (CFI). Z.Y. was supported by the Canada Research Chairs Program.

\appendix

\section{Fourier series of thermoelastic stress calculation}
By combining Eqs. \ref{eq:13} and \ref{eq:14}, we can eliminate the electronic temperature from Eq. \ref{eq:13}, and by introducing $\psi_1 = \frac{i \omega (C_e + C_{\mathrm{lat}})}{\kappa_{e}} - \frac{C_e C_{\mathrm{lat}} \omega^2}{\kappa_{e} G}$, $\psi_2 = 1 + \frac{i \omega C_{\mathrm{lat}}}{G}$, $\psi_3 = \left( \frac{K \beta_T T_0}{G} \right) \cdot \frac{C_e \omega^2 - i \omega G}{\kappa_{e}}$, $\psi_4 = \frac{K \beta_T T_0 i \omega}{G}$, $\gamma = \frac{\beta}{\kappa_{e}}$, $\theta_{e} = T_{e} - T_0$, $\theta_{\mathrm{lat}} = T_{\mathrm{lat}} - T_0$, $c_e = \sqrt{\frac{K + \frac{4 \mu}{3}}{\rho}}$, and $v = -\frac{K \beta_T}{K + \frac{4 \mu}{3}}$, the governing equations can be written as
\begin{equation} \label{eq:20}
	\psi_1 \theta^{\mathrm{lat}}_{i} = \psi_2 \frac{\partial^2 \theta^{\mathrm{lat}}_{i}}{\partial x^2}+\psi_3 \frac{\partial u_i}{\partial x}+\psi_4 \frac{\partial^3 u_i}{\partial x^3}+\gamma e^{-\beta x},
\end{equation}
\begin{equation} \label{eq:21}
	\theta^{e}_{i}=\frac{1}{G}(\left(i \omega (C_{\mathrm{lat}}+G) \theta^{\mathrm{lat}}_i+i \omega K \beta_{T} T_0 \frac{\partial u_i}{\partial x}\right),
\end{equation}
\begin{equation} \label{eq:22}
	-\frac{\omega^2 u_i}{c_{\mathrm{e}}^2}=\frac{\partial^2 u_i}{\partial x^2}+v \frac{\partial \theta^{\mathrm{lat}}_i}{\partial x},
\end{equation}
As Eqs. \ref{eq:20} and \ref{eq:22} are decoupled from the electronic temperature, they can be solved simultaneously.
Once $\theta^{\mathrm{lat}}_i$ and $u_i$ have been obtained, Eq. \ref{eq:21} can be calculated.
The final solutions to Eqs. \ref{eq:20} and \ref{eq:22} are in the form of \cite{wang2001thermoelastic}
\begin{equation} \label{eq:23}
	\theta^{\mathrm{lat}}_i = A_{1, i}e^{k_{1,i}x} + A_{2, i}e^{k_{2,i}x} + A_{p, i}e^{-\beta x}
\end{equation}
\begin{equation} \label{eq:24}
	u_i = -\frac{vk_{1,i}A_{1,i}}{\omega^2/c^2_e+k^2_{1,i}}e^{k_{1,i}x} -\frac{vk_{2,i}A_{2,i}}{\omega^2/c^2_e+k^2_{2,i}}e^{k_{2,i}x} + B_{p, i}e^{-\beta x}
\end{equation}
where $A_{p, i}$ and $B_{p, i}$ are obtained by substituting Eqs. \ref{eq:20} and \ref{eq:22} into the particular solutions $\theta^{\mathrm{lat}}_{p,i} = A^{\mathrm{lat}}_{p,i}e^{-\beta x}$ and $u_{p,i} = B_{p,i}e^{-\beta x}$ and the solutions to $A_{1, i}$ and $A_{2, i}$ are obtained by imposing the boundary condition of $\frac{\partial \theta^{\mathrm{lat}}_i}{\partial x}=0$ and $\frac{\partial u_i}{\partial x}+v \theta^{\mathrm{lat}}_i=0$ on to Eqs. \ref{eq:23} and \ref{eq:24}.
The parameters $k_{1,i}$ and $k_{2,i}$ are calculated as a byproduct of the particular solutions \cite{wang2001thermoelastic}.
Finally, the stress field can be expressed as
\begin{equation}
	\begin{aligned}
		\sigma_i &= \left(K+\frac{4}{3} \mu\right) \cdot \left(\frac{\partial u_i}{\partial x}+v \theta^{\mathrm{lat}}_i\right) \\
		&= (K + \frac{4 \mu}{3})\cdot(e^{k_{1,i}x}(B_{1,i} k_{1,i}+v A_{1,i})+e^{k_{2,i}x}(B_{2,i} k_{2,i}+v A_{2,i})) \\
		& +e^{-\beta x}(-\beta B_{\mathrm{p},i}+v A_{\mathrm{p},i}))
	\end{aligned}
\end{equation}
where the Fourier parameters are given by
\begin{equation}
	A_{p,i} = \frac{(\beta^2 + \omega^2/c^2_e)\gamma}{(\psi_1 - \beta^2\psi_2)(\beta^2 + \omega^2/c^2_e)+v\beta (\psi_3\beta+\psi_4\beta^3)},
\end{equation}
\begin{equation}    
	B_{p,i} =  \frac{\gamma}{(\psi_1 - \beta^2\psi_2)\left(\frac{\beta^2 + \omega^2/c^2_e}{v\beta}\right) + (\psi_3\beta+\psi_4\beta^3)},
\end{equation}
\begin{equation} 
	k_{1,i} = -\sqrt{\frac{\left(\psi_1 + \epsilon - \left(\omega^2/c^2_e\right) \psi_2 + \psi_3 v \right)}{2 \left(\psi_2 - \psi_4 v \right)}},
\end{equation}
\begin{equation}
	k_{2,i} = -\sqrt{\frac{\left(\psi_1 - \epsilon - \left(\omega^2/c^2_e\right) \psi_2 + \psi_3 v \right)}{2 \left(\psi_2 - \psi_4 v \right)}},
\end{equation} 
\begin{equation}
	\resizebox{.9\hsize}{!}{$
		\epsilon = \sqrt{\left(\psi^2_2 \left(\omega^2/c^2_e\right) + 2\left(\omega^2/c^2_e\right)\psi_1 \psi_2 - 4\psi_4\left(\omega^2/c^2_e\right)\psi_1 v - 2\left(\omega^2/c^2_e\right)\psi_2 \psi_3 v + \psi^2_1 + 2 \psi_1 \psi_3 v + \psi^2_3 v^2\right)}$},
\end{equation}
\begin{equation}
	\resizebox{.9\hsize}{!}{$
		A_{1,i} = \frac{{(k^2_{1,i} + \omega^2/c^2_e)\left(A_{p,i}k^3_{2,i}v - B_{p,i}\beta k^3_{2,i} - \left(\omega^2/c^2_e\right) B_{p,i}\beta k_{2,i}  + \left(\omega^2/c^2_e\right) A_{p,i}\beta v  + \left(\omega^2/c^2_e\right) A_{p,i} k_{2,i}v\right)}}{{\left(\omega^2/c^2_e\right) (k_{1,i} - k_{2,i})v(k^2_{1,i} + k_{1,i} k_{2,i} + k^2_{2,i} + \omega^2/c^2_e)}}$},
\end{equation}
\begin{equation}
	\resizebox{.9\hsize}{!}{$
		A_{2,i} = \frac{{(k^2_{2,i} + \omega^2/c^2_e)\left(A_{p,i}k^3_{1,i}v - B_{p,i}\beta k^3_{1,i} - \left(\omega^2/c^2_e\right) B_{p,i}\beta k_{1,i}  + \left(\omega^2/c^2_e\right) A_{p,i}\beta v  + \left(\omega^2/c^2_e\right) A_{p,i} k_{1,i}v\right)}}{{\left(\omega^2/c^2_e\right) (k_{1,i} - k_{2,i}) v(k^2_{1,i} + k_{1,i} k_{2,i} + k^2_{2,i} + \omega^2/c^2_e)}}$},
\end{equation}
\section{Nickel material properties used in analytical calculation}
\begin{table}[h!]
	\small
	\begin{center}
		\caption{The material properties of Ni used in the analytical calculations. Values were obtained from \cite{hohlfeld2000electron, cormier2015finite, bae2008general,kollie1977measurement}.}
		\label{tbl:2}    
		\begin{tabular}{c c}
			\hline
			\hline
			Properties (unit) \\ \hline     \hline
			Density, $\rho$ ($\mathrm{kg} \mathrm{~m}^{-3}$) & 8890\\
			Shear Modulus, $\mu$ (GPa) & 76\\
			Bulk Modulus, $K$ (GPa) & 180\\    
			Thermal conductivity, $\kappa$ ($\mathrm{W} \mathrm{~m}^{-1} \mathrm{~K}^{-1}$) & 91\\
			Volumetric lattice heat capacity, $C_{\mathrm{lat}}$ ($\mathrm{~J} \mathrm{~m}^{-3} \mathrm{~K}^{-1}$) & $3.5 \times 10^{6}$\\
			Volumetric electronic heat capacity, $C_e$ ($\mathrm{~J} \mathrm{~m}^{-3} \mathrm{~K}^{-1}$) & $3.7 \times 10^{5}$\\
			Volumetric thermal expansion coefficient, $\beta_T$ ($\mathrm{~K}^{-1}$) & $1.3 \times 10^{-5}$ \\
			Optical absorption coefficient, $\beta$ ($\mathrm{~m}^{-1}$) & $7.4 \times 10^{7}$ \\
			Electron-phonon coupling strength, $G$ ($\mathrm{Wm}^{-3} \mathrm{~K}^{-1}$) & $3.6 \times 10^{17}$ \\
			Sound velocity, $c_0$ ($\mathrm{m} \mathrm{~s}^{-1}$) & 4650\\
			Slope of $U_s$ versus $U_p$, $s$ & 1.45\\
			Melting temperature, $T_m$ (K) & 1730\\    
			Reference temperature, $T_0$ (K) & 298\\    
			\hline
			\hline
		\end{tabular}    
	\end{center}
\end{table}
 \bibliographystyle{elsarticle-num} 
 \bibliography{cas-refs}

\end{document}